\documentclass[twocolumn]{aastex631}

\shorttitle{Phase-resolved Retrievals of WD-0137B and EPIC-2122B}
\shortauthors{Lothringer et al.}
\begin{document}

\title{Atmospheric Retrievals of the Phase-resolved Spectra of Irradiated Brown Dwarfs WD-0137B and EPIC-2122B}

\author[0000-0003-3667-8633]{Joshua D. Lothringer}
\affiliation{Department of Physics, Utah Valley University, Orem, UT, USA}
\affiliation{Space Telescope Science Institute, Baltimore, MD, USA}

\author[0000-0003-2969-6040]{Yifan Zhou}
\affiliation{Department of Astronomy, University of Virginia, Charlottesville, VA, USA}

\author[0000-0003-3714-5855]{Dániel Apai}
\affiliation{Department of Astronomy and Steward Observatory, The University of Arizona, Tucson, AZ, USA}
\affiliation{Lunar and Planetary Laboratory, The University of Arizona, Tucson, AZ, USA}

\author[0000-0003-2278-6932]{Xianyu Tan}
\affiliation{Tsung-Dao Lee Institute, Shanghai Jiao Tong University, Shanghai, People’s Republic of China}
\affiliation{School of Physics and Astronomy, Shanghai Jiao Tong University, Shanghai, People’s Republic of China}

\author[0000-0001-9521-6258]{Vivien Parmentier}
\affiliation{Universit\'{e} C\^{o}te d’Azur, Observatoire de la C\^{o}te d’Azur, CNRS, Laboratoire Lagrange, Nice 06304, France}

\author[0000-0003-2478-0120]{Sarah L. Casewell}
\affiliation{School of Physics and Astronomy, University of Leicester, Leicester, UK}



\begin{abstract}

We present an atmospheric retrieval analysis of HST/WFC3/G141 spectroscopic phase curve observations of two brown dwarfs, WD-0137B and EPIC-2122B, in ultra-short period orbits around white dwarf hosts. These systems are analogous to hot and ultra-hot Jupiter systems, enabling a unique and high-precision comparison to exoplanet systems. We use the PETRA retrieval suite to test various analysis setups, {including joint-phase retrievals, multiple temperature structures, and non-uniform abundances}. We find that WD-0137B has a dayside that closely resembles that of other ultra-hot Jupiters with inverted temperature structures and H$^-$ opacity, but quickly transitions to a mostly non-inverted temperature structure on the nightside. Meanwhile, EPIC-2122B's atmosphere remains inverted at all constrained longitudes, with dominant H$^-$ opacity. Retrievals with multiple temperature profiles and non-uniform vertical abundances were generally not statistically justified for this dataset, but retrievals with dayside-dilution factors were found to be justified. Retrieving all phases simultaneously with a linear combination of a dayside and nightside atmosphere was found to be an adequate representation of the entire phase-curve once a longitudinal temperature gradient free parameter was included in the retrieval. {Comparing to global circulation models, we attribute behavior in the 1D retrievals to the inclined viewing geometry of the systems, which results in always-visible irradiated and inverted portions of the atmosphere ``contaminating" spectra measured from the nightside hemisphere. This study sheds light on the similarities between these irradiated brown dwarf systems and hot and ultra-hot Jupiters, but also their unique differences, including the influence of the inclined viewing geometry.}



\end{abstract}

\received{2024 February 9}

\accepted{2024 April 24}



\section{Introduction} \label{sec:intro}

While the vast majority of brown dwarfs are either isolated field dwarfs or wide-separation companions \citep{dupuy:2012}, a valuable handful of brown dwarfs are found to be highly irradiated by a close companion. A small number of irradiated brown dwarfs are found near main-sequence stars, such as KELT-1b \citep{siverd:2012,beatty:2017b} or CWW 89Ab \citep{curtis:2016,nowak:2017,beatty:2018b}, but there also exists a population of brown dwarfs orbiting close enough to white dwarfs that they can reach equilibrium temperatures of hundreds to thousands of Kelvin \citep[e.g.,][]{farihi:2004,burleigh:2006}. These white-dwarf-brown-dwarf (WD-BD) systems provide a unique comparison to exoplanets around main-sequence stars.

These can orbit less than a solar radius from their WD primary, resulting in very short orbital periods on the order of hours. While the shortest {gas giant} orbital period, TOI-2109, is 16.14 hours \citep{wong:2021:2109}, the shortest BD-WD orbital period, EPIC-2122B, is only 68 minutes \citep{casewell:2018}. Additionally, the brown dwarf companions are significantly larger than their WD primary, leading to relatively large flux ratios.{ While hot and ultra-hot Jupiters can have flux ratios of thousands of parts per million \citep[e.g.,][]{coulombe:2023}, brown dwarf companions can actually exceed the flux from their WD primary in the infrared (i.e., a flux ratio of $>100\%$).} BD-WD systems thus offer valuable opportunities to observe high signal-to-noise phase curves \citep[e.g.,][]{zhou:2022,lew:2022,amaro:2023}.

The entire spectroscopic phase curves of two of these BD-WD systems, WD-0137B and EPIC-2122B, have recently been observed by the \textit{Hubble Space Telescope} (HST) \citep{zhou:2022}. While WD-0137B is more similar to the average hot Jupiter in the level of irradiation, EPIC-2122B is comparable to the hottest Jovian systems, called ultra-hot Jupiters \citep{parmentier:2018,lothringer:2018b,kitzmann:2018}. Both objects are non-eclipsing with orbital inclinations of 35$^{\circ}$ for WD-0137B \citep{maxted:2006} and 56$^{\circ}$ for EPIC 2122B \citep{casewell:2018}. This means that the nadir point (i.e., the point directly facing Earth) on the brown dwarfs are at latitudes of 55$^{\circ}$ and 34$^{\circ}$ respectively, providing a unique opportunity to understand the mid-latitude regions of irradiated atmospheres. 

\cite{zhou:2022} found WD-0137B to have clear evidence of water absorption on the nightside of the brown dwarf, while the dayside is more featureless, indicative of either H$_2$O dissociation or an isothermal atmosphere. EPIC-2122B, on the other hand, hardly shows any H$_2$O absorption at all. These spectra were fit with a grid of forward models, expanded from \citep{lothringer:2020c}, alongside global-circulation models \citep{tan:2019b,tan:2020}. These self-consistent models are limited in their flexibility to fit the observations and cannot robustly quantify our understanding and uncertainty in atmospheric properties like the temperature structure and molecular abundances.

In this work, we extend this analysis to include a suite of data-driven atmospheric retrievals using the PHOENIX Exoplanet Retrieval Algorithm (PETRA)  framework \citep{lothringer:2020a}. This modeling provides a flexible framework through which the observations themselves guide the fit to the models, allowing for a more robust statistical determination of the atmospheric properties. After presenting the analyses, we compare the results to those from the self-consistent forward models and the global circulation model for each of the systems.

\section{Methods}

\subsection{Observations}\label{observations}

We use the data first analyzed and described by \cite{zhou:2022}. WD-0137B and EPIC-2122B were each observed with the \textit{Hubble Space Telescope} (HST) using the Wide Field Camera 3 (WFC3)'s G141 grism ($\approx$1.1-1.7~$\mu$m) as part of HST Program 15947 (PI: Apai, \cite{apai:15947}). WD-0137B was observed over the course of five HST orbits from UTC 2020 June 20 18:46:25 to 2020 June 21 01:50:02, accumulating 70 spectra that covered 4.1 orbital periods of the brown dwarf. EPIC-2122B was similarly observed over the course of four HST orbits from UTC 2020 May 06 18:09:08 to 2020 May 06 23:35:11, accumulating 384 spectra covering 5.6 orbital periods of the brown dwarf.

\begin{figure}[t]
\epsscale{1.15}
\plotone{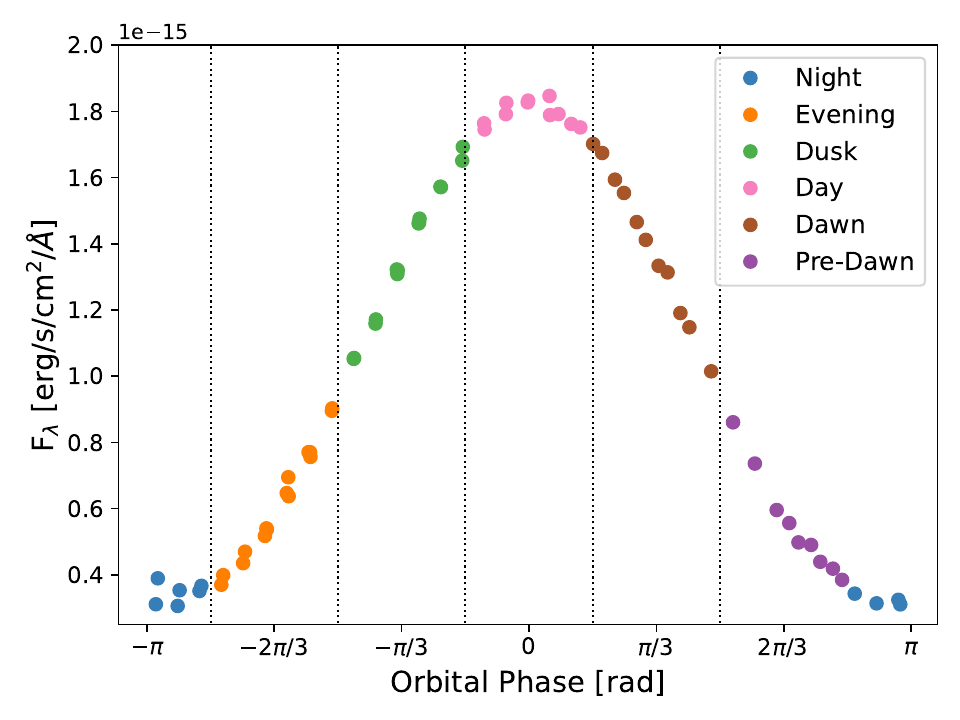}
\plotone{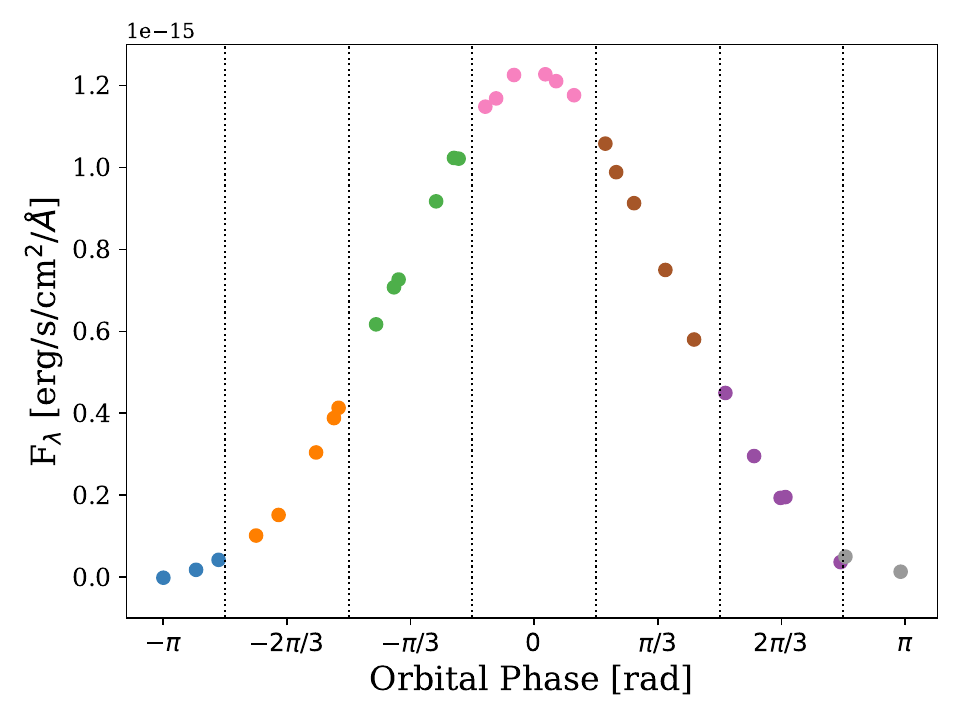}
\caption{White-light phase curve of WD-0137B (top) and EPIC-2122B (bottom) as analyzed by \cite{zhou:2022}. The phase curve is divided into six equal parts as a function or orbital phase, corresponding to night (blue), evening (orange), dusk (green), day (pink), dawn (brown), and pre-dawn (purple), from left to right. Uncertainties on the total flux for each spectrum are smaller than the dot size.
\label{fig:WLC_PC}}
\end{figure}

After initial data reduction, the observations were corrected for systematics related to charge-trapping on the detector using the RECTE model of \cite{zhou:2017}. {The white dwarf component was subtracted using a white dwarf atmosphere model \citep{koester:2010} convolved with a Gaussian kernel representing instrument broadening (R=130 at 1.4~$\mu$m). The Pa~$\beta$ absorption line is the only noticeable spectral feature in the WD spectrum, is well-fit by the WD model, and does not appear in the residuals. In total, uncertainty in the WD spectra contributes to 49\% and 6.7\% of the uncertainty on the BD spectrum for WD-0137B and EPIC-2122B, respectively}. We refer the reader to \cite{zhou:2022} for further details regarding the data analysis. A spectroscopic time-series of the brown dwarf at the native resolving power of R$\sim$130 (at 1.4~$\mu$m) was then formed, which was further binned into six bins based on the orbital phase of the binary, as was done \cite{zhou:2022}. The phase-bins were centered at $-\frac{2}{3}\pi$, $-\frac{1}{3}\pi$, 0, $\frac{1}{3}\pi$, $\frac{2}{3}\pi$, and $\pi$ radians, shown for WD-0137B and EPIC-2122B in Figure~\ref{fig:WLC_PC}. We term these phases ``Pre-Dawn", ``Dawn", ``Day", ``Dusk", ``Evening", and ``Night", respectively. The spectra at each of these phases for WD-0137B and EPIC-2122B are shown in Figures~\ref{fig:wd0137_fits} and~\ref{fig:EPIC2122_fits}, respectively.

\begin{table*}
\centering
\caption{Relevant System Parameters \label{tab:properties}}
\begin{tabular}{c||c|c|c}
\hline
 \hline
 Property & WD-0137$^{a,b}$ & EPIC-2122$^c$ & WASP-121$^d$ (comparison) \\
 \hline
 Primary T$_{eff}$ (K)           & 16500 $\pm 500$  & 24490 $\pm 194$ & 6459 $\pm 140$   \\
 Primary Radius (R$_{\sun}$)  & 0.0186 $\pm$ 0.0012  & 0.017 $\pm 0.005$ & 1.458 $\pm 0.030$  \\
 Primary log(g) ($cgs$)  & 7.49 $\pm$ 0.08  & 7.63 $\pm$0.02 & 4.242 $^{+0.011}_{-0.012}$  \\
  Orbital Separation  (R$_{\sun}$)      & 0.65 $\pm$ 0.025 & 0.44 {$\pm$ 0.03} & 5.47 $^{+0.106}_{-0.108}$  \\
  Orbital Period (min)     & 114.3729 $\pm$ 4e-5 & 68.21218 $\pm$ 8e-5 & 1,835.8927  $^{+0.00029}_{-0.00036}$ \\
Secondary T$_{eq}$ (K) & 1970 $\pm$ 60  & 3404 $\pm$ 30 & 2358 $\pm 52$  \\
Secondary Mass    ($M_{J}$)    & 55.5 $\pm 6.31$  & 58$^{+7}_{-11}$  & 1.183$^{+0.064}_{-0.062}$  \\
Secondary Radius$^e$  ($R_{J}$)      & 0.973  & 0.947 & 1.865 $\pm0.044$  \\
Secondary log(g)   ($cgs$)          & 5.18$\pm$ 0.03 & 5.2 $^{+0.05}_{-0.09}$ & 2.97   \\
 \hline
\end{tabular}
    \tablenotetext{}{a) \cite{maxted:2006}, b) \cite{casewell:2015}, c) \cite{casewell:2018}, d) {\cite{delrez:2016}, e) estimated in a) and c) based on the mass and age.} }

\end{table*}

\subsection{Retrieval Framework}\label{sec:methods:retrieval}

We use the PHOENIX Exoplanet Retrieval Algorithm (PETRA) \citep{lothringer:2020a}, which employs the PHOENIX atmosphere model \citep{hauschildt:1999,barman:2001} as the forward model in a MCMC-based statistical retrieval framework. While initially developed for work on exoplanet atmospheres, the flexibility of the PHOENIX atmosphere model enables application to a wide variety of objects, including brown dwarfs. We previously used PHOENIX to calculate self-consistent models of the brown dwarf's highly irradiated atmosphere. Here, we use it in the retrieval framework, providing a direct comparison to the analysis of \cite{lothringer:2020c} and \cite{zhou:2022}. 


We refer the reader to \cite{lothringer:2020a} for more information regarding PETRA, but we summarize our retrieval setup here. In all analyses, we use the physically-motivated radiative-equilibrium temperature structure parameterization of \cite{parmentier:2014}, which uses five parameters to describe the full temperature structure. This temperature structure is flexible enough to allow for temperature inversions, which is important in these highly irradiated objects \citep{lothringer:2020a}. We fixed the internal temperature for each object to the best-fit value from a {fiducial} retrieval (described below) of the non-irradiated nightside phases where the internal temperature was allowed as a free parameter, finding 1,520 $^{+30}_{-130}$~K for WD-0137B and 1,570$^{+225}_{-130}$~K for EPIC-2122B. {This retrieval had the flexibility to partition the nightside temperature between the irradiation/advected temperature, through $\beta$ in the parameterization, and the temperature of the deeper atmosphere, through T$_{int}$.}
This was important for these objects because brown dwarfs can have large internal temperatures (i.e., their $T_{\rm{eff}}$ if they were non-irradiated) and it is unknown \textit{a priori} how much of an irradiated brown dwarf's temperature structure is driven by its internal heat versus the irradiation. {Note however, that within the \cite{parmentier:2014} parameterization, we also fit for $\kappa_{ir}$, which maps the temperature-optical depth profile onto a pressure grid. This means that while the deep temperature-optical depth profile will match between models with the same internal temperature, their deep temperature-pressure profile will not necessarily match.}

{We fixed the brown dwarfs' surface gravity to literature values (see Table~\ref{tab:properties} for object system parameters)}, which come from orbital fits and will generally be more constraining ($\pm$0.05 and 0.1, respectively) than values fit from the retrieval at the relatively limited wavelength coverage and low spectral-resolution of the observations, while also preventing potential degeneracies in the retrieval. For the latter point, surface gravity varies the vertical position of the temperature structure (i.e., the pressure at the photosphere). This behavior is encapsulated in the $\kappa_{IR}$ free parameter in the \cite{parmentier:2014} parameterization, which maps the optical-depth, $\tau$, and pressure, $P$, via $\kappa_{IR}=\tau P/g$. Including both $\kappa_{IR}$ and $g$ would result in a strong degeneracy in the retrieval. 

However, PHOENIX uses the optical depth as its main vertical coordinate and calculates it self-consistently from our assumed temperature and pressure. Hydrostatic balance, and subsequently the radius as a function of pressure, is calculated with this self-consistent optical depth and assumed surface gravity. Thus, the surface gravity has an additional effect on the vertical structure of the atmosphere beyond $\kappa_{IR}$. So, while we fix the surface gravity to literature values to avoid this degeneracy, the vertical position of the temperature structure and the pressure of the photosphere in the retrieved results should be interpreted with this knowledge.

{Our fiducial retrieval approach was a single 5-parameter temperature-pressure (TP) profile with uniform vertical abundances fit to each phase independently.} We also computed retrievals that take into account the non-uniform nature of these brown dwarfs' hemispheres as observed during our observations. In particular, global circulation models have previously indicated that the dayside hemispheres can exhibit irradiation-driven hotspots at the subsolar point that are physically smaller than the greater hemisphere but have a significantly higher temperature \citep{lee:2020,tan:2020}. Similarly, the orbital inclination of WD-0137B and EPIC-2122B means that portions of the high-latitude but still irradiated dayside hemisphere will be visible at nightside phases and vice-versa. These hotspots and hemisphere regions can end up dominating or diluting the outgoing flux. 

We therefore ran retrievals with a multiplicative filling factor parameter as described by \cite{taylor:2020} that can account for a small hot spot dominating the observed spectra. We also experimented with retrievals that fit for two independent temperature structures on the dayside hemisphere, similar to the ``2TP-Fixed/Free" scenarios from \cite{feng:2020}. Abundances, described below, were also allowed to be independently fit to each structure in this setup. The fluxes from these two structures were then linearly combined according to a filling factor parameter:
\begin{equation}\label{eq:phi}
	f_{t} = \eta*f_{d} + (1-\eta)*f_{n},
\end{equation}
where $f_{t}$ is the total flux, $f_{d}$ is the dayside flux, $f_{n}$ is the nightside flux, and $\eta$ is the filling factor parameter.

Because we focus on the HST/WFC3/G141 observations (see Section~\ref{observations}), we restrict our opacity calculation to species that absorb in the relevant wavelength range (1.1-1.7$\mu$m). H$_2$O is the main absorber in this region and we use the high-temperature BT2 line lists \citep{barber:2006}. Studies of ultra-hot Jupiter exoplanets have revealed the importance of H$^-$ opacity over this wavelength range in atmospheres $\gtrsim$2000~K \citep{arcangeli:2018,parmentier:2018,lothringer:2018b,kitzmann:2018}. We therefore include H$^-$ opacity using the equations of \cite{john:1988}. In the retrievals, we vary H$_2$O and H$^-$ (via the e$^-$ density) as free parameters to fit to the observations. Our fiducial retrievals assume uniform vertical abundances, which is the simplest case and is often assumed in free retrievals of exoplanets \citep[e.g.,][]{madhusudhan:2009}. 

However, we also explore retrievals with non-uniform abundances using the parameterization of \cite{lothringer:2020a}. The abundance of both H$_2$O and H$^-$ are not expected to necessarily be vertically uniform in the highly-irradiated objects in the present study. For H$_2$O, the high temperatures, particularly in the dayside temperature inversion, can thermally dissociate the molecule, resulting in an abundance that quickly drops with altitude \citep{lothringer:2018b,parmentier:2018}. On the other hand, H$^-$ depends sensitively on the abundance of both atomic H and the e$^-$ density, which themselves do not have uniform vertical abundances. e$^-$ density will increase in the temperature inversion as thermal ionization increases, while H$_2$ will thermally dissociate to atomic H beginning near the base of the temperature inversion, causing a sudden increase in the atomic H abundances with altitude.

In our fiducial retrieval, each of the six orbital phases is fit independently of the others. However, we also explored a joint retrieval where the six phases were fit simultaneously with two different ``atmospheres", nominally an atmosphere representative of the dayside and an atmosphere representative of the nightside. While the temperature structures and molecular abundances of the two hemispheres were completely independent (similar to the dual temperature structure described above), they were related to planetary flux via a phase-dependent version of Equation~\ref{eq:phi} that represents the amount of the dayside versus nightside hemisphere seen at each phase angle \citep{feng:2020}:
\begin{equation}\label{eq:etaphi}
	\eta_{\phi} = [1+\cos(2\pi\phi-c)]/2 
\end{equation}
where $\phi$ is the orbital phase and $c$ is a phase offset parameter to account for the advection of any hot spot from the substellar point. In this framework, we can use all of the phase curve information simultaneously to inform the fit. We found it necessary to add an additional free parameter, $\delta$, to account for the magnitude of the day-night gradient, such that Equation~\ref{eq:etaphi} becomes:
\begin{equation}\label{eq:etadelta}
	\eta_{\phi} = [1+\cos(2\pi\phi-c)^{\delta}]/2. 
\end{equation}
A similar effect could be obtained by including higher-order Fourier series terms as done in the phase-curve fits in \cite{zhou:2022}, but to reduce the number of parameters, we include a single parameter for the power of the cosine.

Each retrieval analysis was run with 1\AA~direct opacity sampling in wavelength across the 1.1-1.7$\mu$m range of the observations. The spectrum was then convolved down to the instrument resolution for comparison to the observations. Retrievals were run with varying numbers of iterations, but generally tens of thousands iterations led to Gelman-Rubin statistics of less than 1.01. {The Gelman-Rubin statistic compares the average variance of each chain across all the chains to the variance of the average of the individual chains, and statistics closer to 1.0 can be considered more converged \citep{gelman:1992}.}

\section{Results}

\subsection{WD-0137B}\label{sec:wd0137}

\begin{figure*}[t]
\plotone{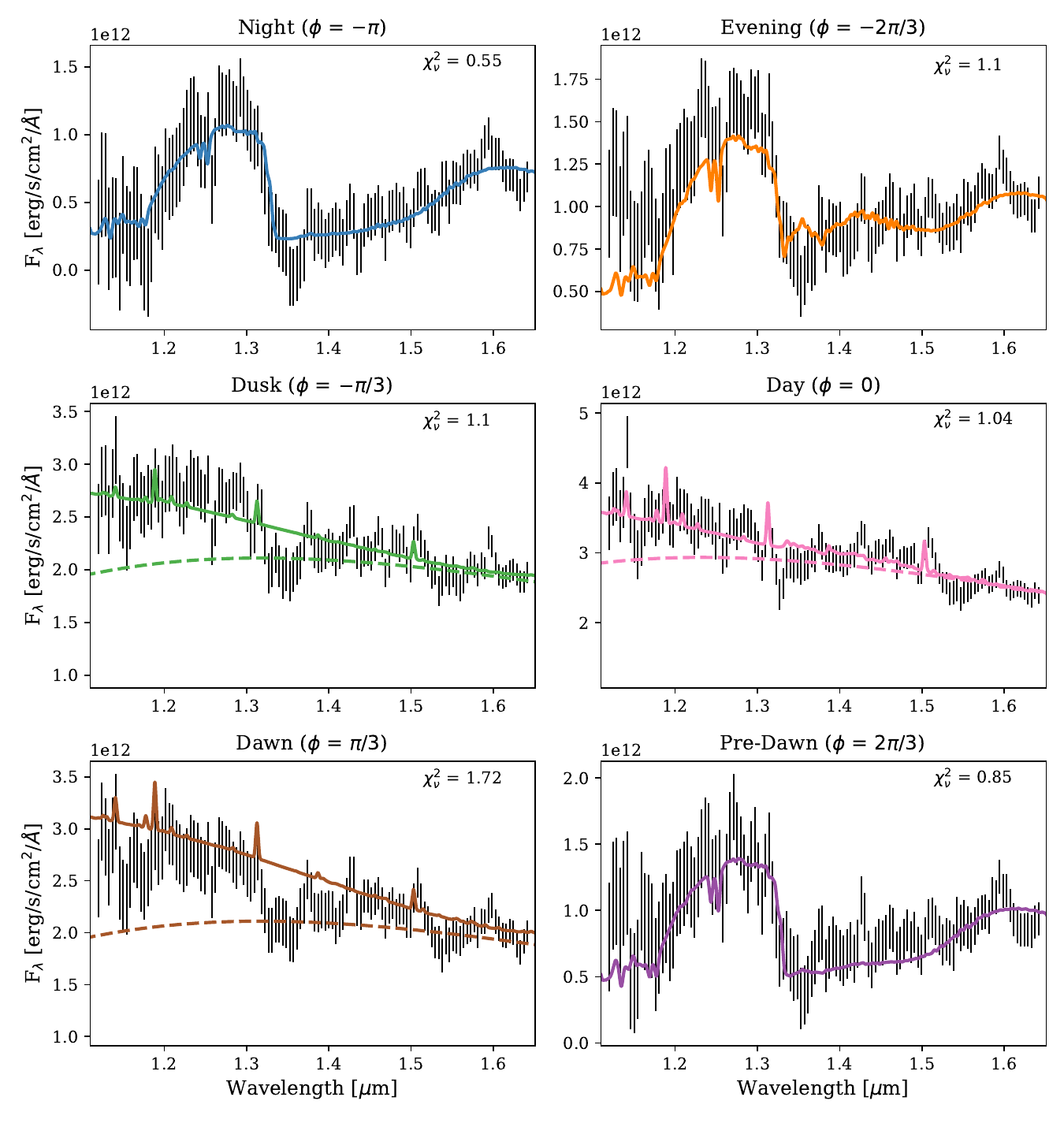}
\caption{Best-fit spectra of WD-0137B from the fiducial retrievals to each of the six hemispheres. The dawn, dusk, and dayside phases also include, respectively, a 2,200, 2,200, and 2,350~K blackbody (dashed line) for reference to show the observed spectrum does not follow a simple blackbody.
\label{fig:wd0137_fits}}
\end{figure*}

Fiducial fits (a single, independent 5-parameter TP-profile, with uniform vertical abundances) to the six phase-resolved spectra of WD-0137B are shown in Figure~\ref{fig:wd0137_fits}. The spectra are well-fit in a $\chi^2$ sense, with a reduced $\chi^2$, $\chi^2_\nu$, being near or below 1.0 with 122 data points and 7 free parameters. The fact $\chi^2_\nu$ falls below 1.0 at some phases suggests the observational uncertainties are overestimated and the  uncertainties on subsequent constraints on the atmosphere properties are similarly overestimated.

\begin{figure}[t]
\epsscale{1.15}
\plotone{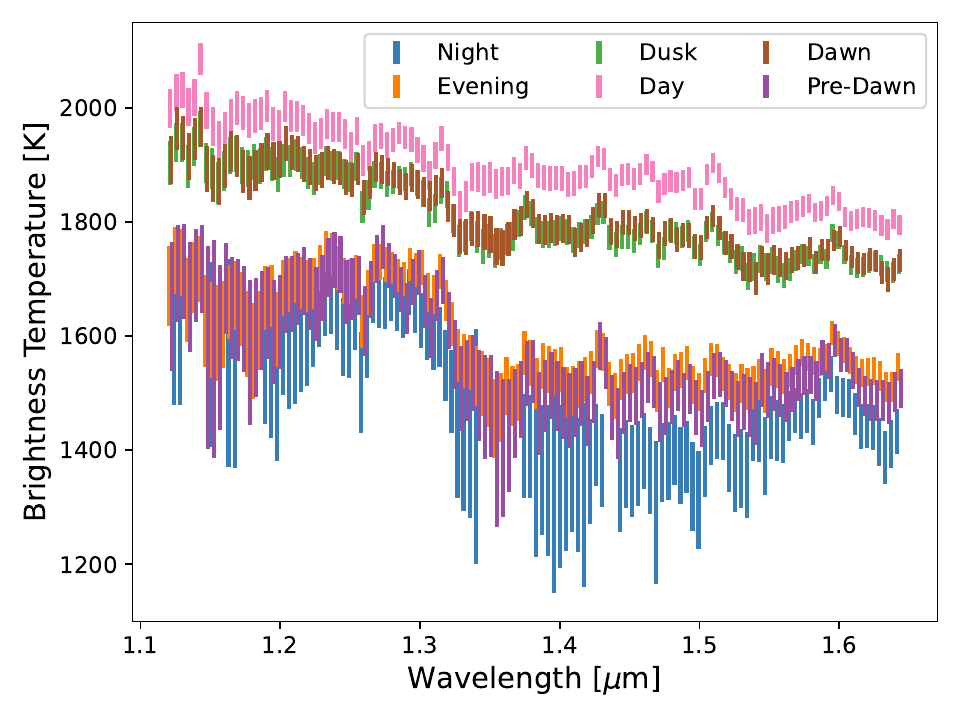}
\caption{Observed brightness temperatures of WD-0137B as a function of wavelength for each of the six hemispheres.
\label{fig:wd0137_BTs}}
\end{figure}

As highlighted by \cite{zhou:2022}, H$_2$O absorption is apparent at nightside orbital phases, however this absorption decreases on the dayside hemisphere. While there is still some residual structure in the dayside spectra (including a consistent dip at about 1.33~$\mu{m}$), the retrieved spectra are relatively featureless. However, these otherwise featureless dayside spectra are not simple blackbodies (dashed line in Figure~\ref{fig:wd0137_fits}), which would be the case if the atmosphere were isothermal. The dayside spectra are in fact ``bluer" than a corresponding blackbody. This is further demonstrated in Figure~\ref{fig:wd0137_BTs}, which shows that the brightness temperatures on the dayside phases decreases with wavelength.

Rather, the spectrum is characteristic of an inverted temperature profile with continuous opacity from H$^-$ \citep{arcangeli:2018, parmentier:2018,lothringer:2018b}. The bound-free opacity of H$^-$ decreases redward toward a minimum at 1.6~$\mu$m (beyond which free-free opacity increases redward). If the temperature profile is inverted (i.e., temperature increases with decreasing pressure) then there will be greater flux shortward of 1.6~$\mu$m relative to the blackbody that matches the 1.6~$\mu$m flux (in this case $\sim$2,350~K). We note that this inference of an inverted dayside atmosphere is supported by the ground-based observation of atomic lines in emission from WD-0137B \citep{longstaff:2017}.


\subsubsection{Temperature Structures}

\begin{figure}[t]
\epsscale{1.15}
\plotone{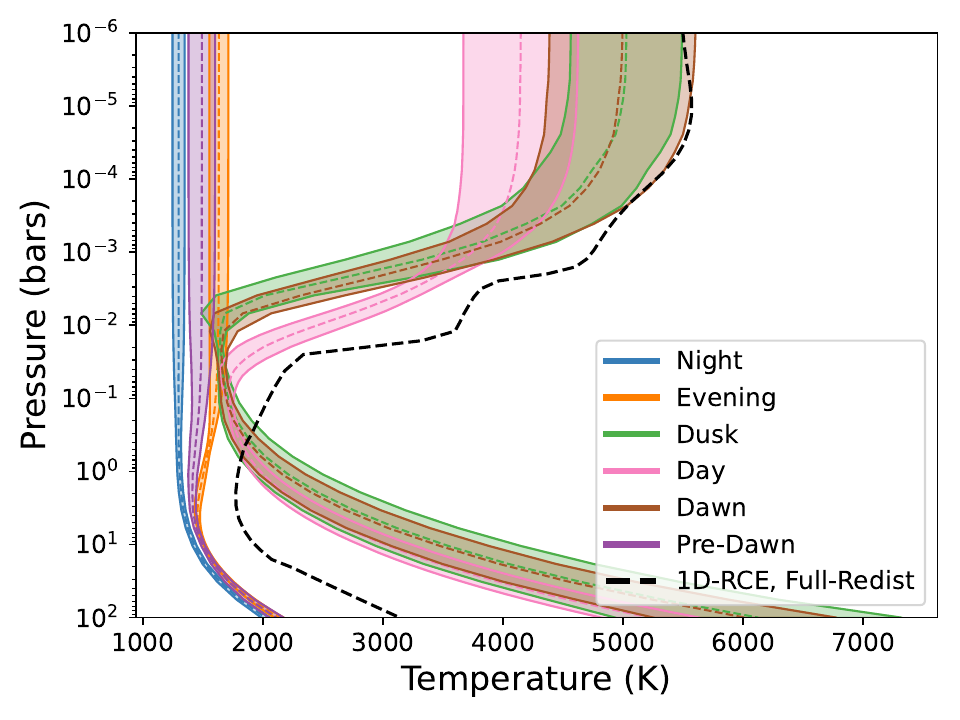}
\caption{Retrieved temperature structure constraints for WD-0137B for each of the six hemispheres, showing the temperature inversion appear on the dayside, weaken, and then disappear as you go toward the nightside. The black dashed line shows the temperature structure from a full-heat redistribution 1D radiative-convective equilibrium model. 
\label{fig:wd0137_TPs}}
\end{figure}

In agreement with the qualitative behavior of the spectra and brightness temperatures, the fiducial retrieved temperature structures are inverted on the dayside and non-inverted on the nightside (see Figure~\ref{fig:wd0137_TPs}). Beyond the inversion, the other major feature of the retrieved temperature structures is the sharp dichotomy between the structures of the dayside and nightside with respect to the deep atmosphere. The nightside phases appear to result in a structure with {the photosphere} shifted towards higher pressure. {For the dayside temperature structures, the photosphere is between 0.1 and 10 bars, while on the nightside the photosphere is generally between 1 and 100 bars.} One reason this might be is the dichotomy between the presence of H$^-$ on the day- and night-side (see Section~\ref{sec:wd0137:abunds}, below). An effect of the strong continuous opacity of H$^-$ is to effectively raise the photosphere to lower pressures \citep{parmentier:2018}, as seen in the dayside retrievals in Figure~\ref{fig:wd0137_TPs}). Similarly, we will see deeper into the atmosphere as the temperature decreases onto the nightside as the H$_2$O opacity outside of the major 1.4~$\mu{m}$ band significantly weakens \citep{tinetti:2012}.

The retrieved dayside temperature structures roughly match the theoretical expectations from a 1D radiative-convective equilibrium models assuming full heat redistribution \citep{lothringer:2020c}. The major difference between the retrieved and theoretical dayside structures is that the entire structure appears offset in pressure, i.e., the photosphere is at a higher altitude in the retrieved structures compared to the theoretical structure. If this difference is indeed physical, it could point to a lower surface gravity than was used to calculate the modeled spectrum (log10($g$)$_{cgs}$=5.1), a higher metallicity, or just a higher abundance of H$^-$. For the latter scenario, photodissociation and photoionization may raise both the abundance of H and e$^-$ in the atmosphere, serving to increase the number of H$^-$ ions. 

\begin{table*}
\centering
\caption{WD-0137B dayside retrieval scenario comparison.\label{tab:wd0137:scenarios}}
 \begin{tabular}{ c||c|c|c|c|c}

\hline
 \hline
 Scenario & N$_{params}$ & -log($\mathcal{L}$) & $\Delta$BIC & $\Delta$AIC & Conclusion \\
 \hline
 Fiducial           & 7  & -3319.9 & --   & --    &   --              \\
 Non-uniform e$^-$  & 9  & -3319.9 & -9.6 & -4.0  &  Not Justified    \\
 Non-uniform H$_2$O & 9  & -3319.1 & -8.0 & -2.4  &  Not Justified    \\
 Dual-TP            & 15 & -3319.9 & -38  & -16   &  Not Justified    \\
 Fill Factor        & 8  & -3316.6 & 1.8  & 4.6   &  Positive evidence\\
 \hline
\end{tabular}
\end{table*}

\subsubsection{Abundances}\label{sec:wd0137:abunds}

\begin{figure*}[ht!]
\epsscale{1.15}
\plottwo{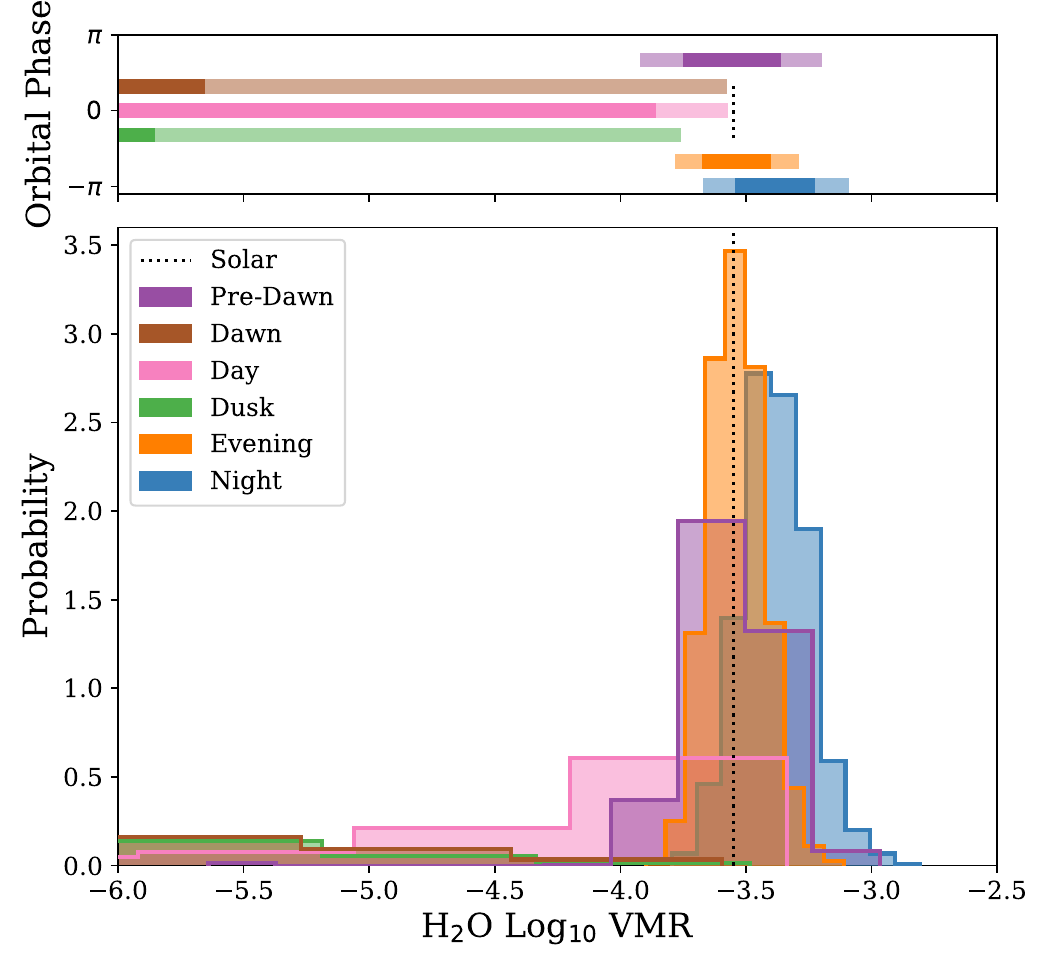}{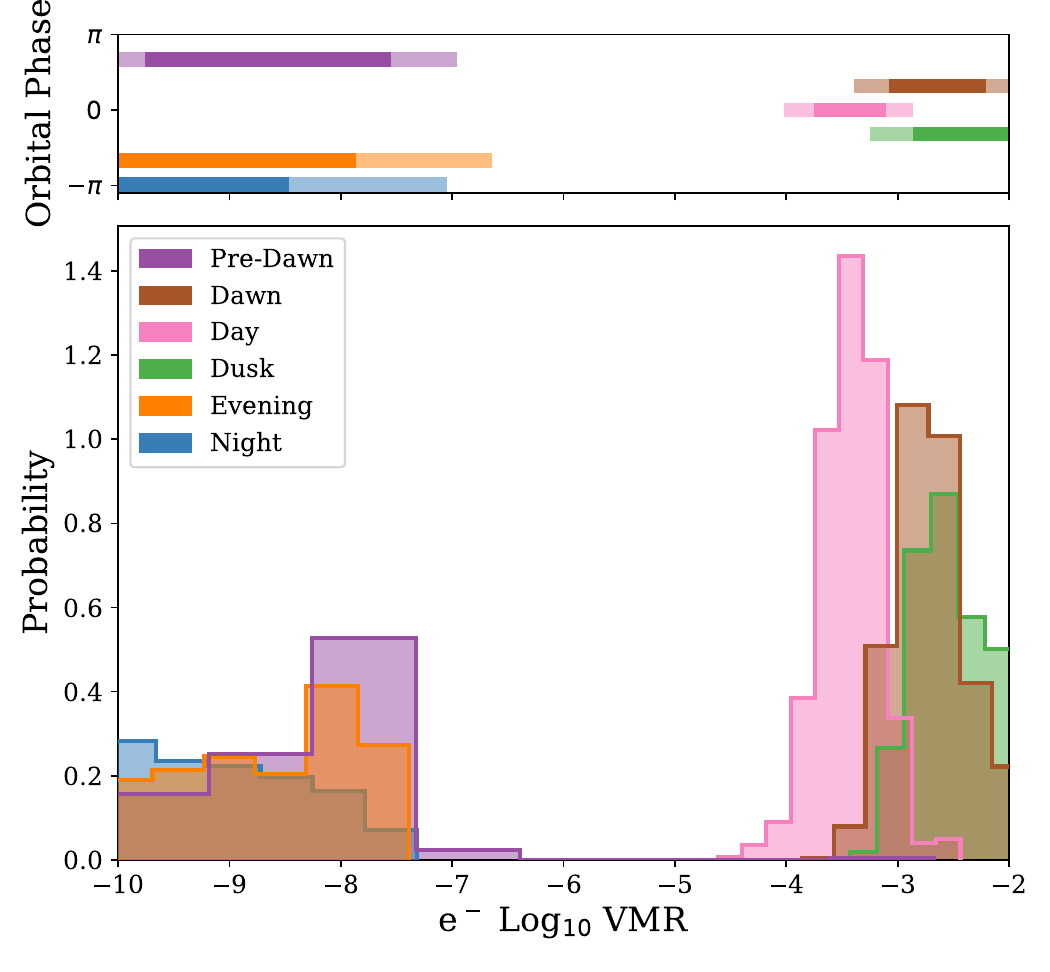}
\caption{Abundance of H$_2$O (top) and e$^-$ in WD-0137B for each of the six hemispheres. H$_2$O is constrained on the nightside hemisphere, but unconstrained on the dayside hemisphere, due to thermal dissociation and the increased opacity of H$^-$, as evidenced by the abundance of e$^-$ on the dayside.
\label{fig:wd0137_abunds}
}
\end{figure*}

Figure~\ref{fig:wd0137_abunds} shows the abundance posterior constraints for H$_2$O and e$^-$ (our proxy for H$^-$) for each of the independently-retrieved phases. H$_2$O was detected at solar abundance on the nightside hemisphere (i.e., evening, night, and pre-dawn sextants), but was undetected on the dayside hemisphere (i.e., day, morning, and dusk sextants). The opposite was found to be true for H$^-$, which was found in abundance on the dayside hemisphere, but was undetected on the nightside hemisphere. These results are consistent with our understanding of highly-irradiated atmospheres \citep[e.g.,][]{parmentier:2018,lothringer:2018b}, which will be hot enough to thermally dissociate H$_2$O on the hot dayside and whose spectra will be further muted by the strong opacity from H$^-$ formed from the high abundance of thermally dissociated molecular H$_2$ and thermally ionized metals. On the cooler nightside, where H$_2$O and H$_2$ recombine, the strong opacity of H$^-$ is absent and the 1.15 and 1.4$\mu$m H$_2$O absorption features are strongly detected. These results suggest that the recombination of H$_2$O and H$_2$ proceeds quickly and in equilibrium, also implying that molecular abundances are not horizontally quenched.

\subsubsection{Non-Fiducial Scenarios}

We explored retrievals with non-uniform vertical e$^-$ and H$_2$O abundances. As described in Section~\ref{sec:methods:retrieval}, the parameterizations for a non-uniform abundance require two extra parameters that define the slope of the non-uniformity and the pressure at which the non-uniformity begins. To quantify whether the inclusion of these two extra parameters was statistically justified, we calculated the change in the Bayesian information criterion, $\Delta$BIC \citep{schwarz:1978}, and the change in the Akaike Information Criterion, $\Delta$AIC \citep{akaike:1974}, between each of these scenarios and the fiducial retrieval using the dayside spectrum of WD-0137B, where we expect the highest amount of H$_2$O dissociation and e$^-$ abundance. As listed in Table~\ref{tab:wd0137:scenarios}, the change in log-likelihood for the non-uniform e$^-$ and H$_2$O scenarios is 0.0 and 0.8, respectively, leading to $\Delta$BIC values of -9.6 and -8.0, indicating that these additional parameters are not justified by the data (i.e., these parameters do not adequately improve the fit to the data as to be justify their inclusion).

In a similar fashion, we explored retrieval scenarios to account for the non-homogenity of the spectra given the fact that the emitting disk of the planet is certainly non-uniform in temperature. We first fit a linear combination of spectra, each from independent TP-profiles that have their own molecular abundances. This scenario doubles the number of parameters plus an additional parameter to determine the proportion for which to combine the two structures and thus results in a large penalty in the $\Delta$BIC for the increased number of free-parameters. With no detected increase in the log-likelihood, this scenario is not justified by the data.

A simpler way to account for the non-homogeneity of the observed hemisphere is to add a dilution or filling factor, as described in Section~\ref{sec:methods:retrieval}. This corresponds to a scenario where a small hot spot dominates the observed flux. For the retrievals of WD-0137B's dayside, this scenario did improve the log-likelihood by about 3.3{ with a $\Delta$BIC of only 1.8}, corresponding to positive but not strong evidence for its inclusion in the retrievals. Unlike in \cite{zhou:2022} where a filling factor of 0.43 was found for the fit to the dayside with self-consistent 1D radiative-convective equilibrium models, the retrieval has enough flexibility in the abundances and temperature profile to adequately fit the dayside spectrum without a filling factor.


\subsubsection{Joint Phase Curve Fit}\label{sec:wd0137:pcfit}

In addition to running retrievals on each phase independently, we fit for each of the 6 phases jointly using combinations of a "dayside" and "nightside" spectrum, as described in Section~\ref{sec:methods:retrieval}. This allows us to fit for the entire data set with far fewer parameters than is required for each phase independently, while also measuring the hotspot offset and day-night gradient. For WD-0137B, we found a small, $<$$3\sigma$ {substellar }hot spot offset of $c = 0.014$ $\pm$ 0.005{ = 0.80 $\pm$ 0.29$^{\circ}$}, while the day-night gradient was found to be represented by $\delta$ = 1.28 $\pm$ 0.02 (see Equation~\ref{eq:etadelta}). This indicates a steeper-than-sinusoidal day-night gradient, where a sinusoidal day-night transition would correspond to $\delta = 1$. This is consistent with the Fourier analysis of \cite{zhou:2022}, which found that a single first-order sinusoid was insufficient to fit the shape of both WD-137B and EPIC-2122B's phase curve.  

\begin{figure*}[ht!]
\epsscale{1.1}
\plottwo{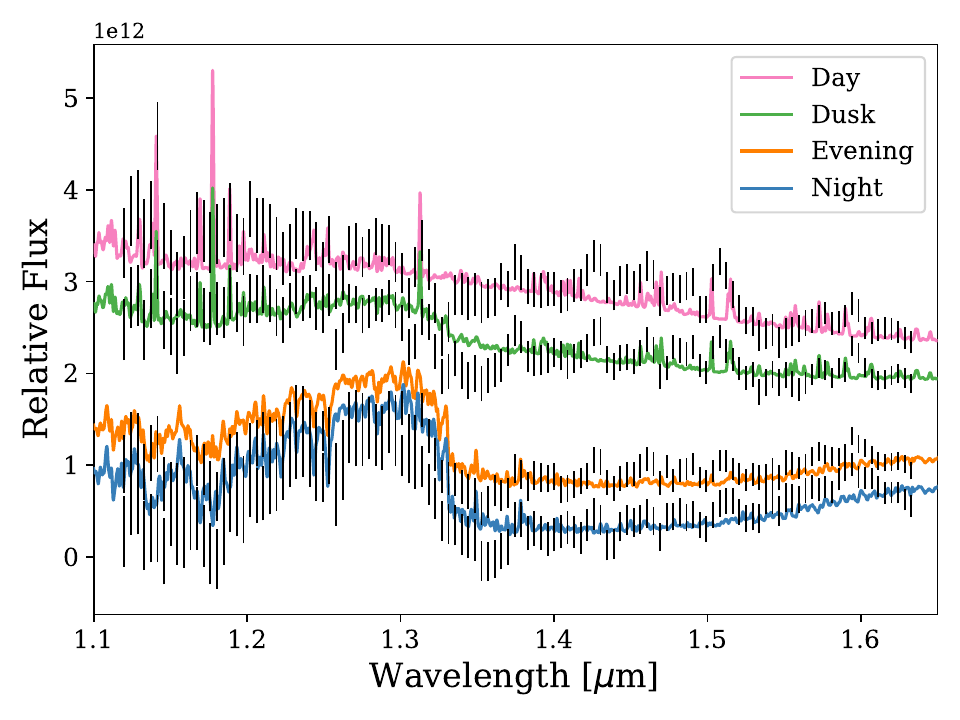}{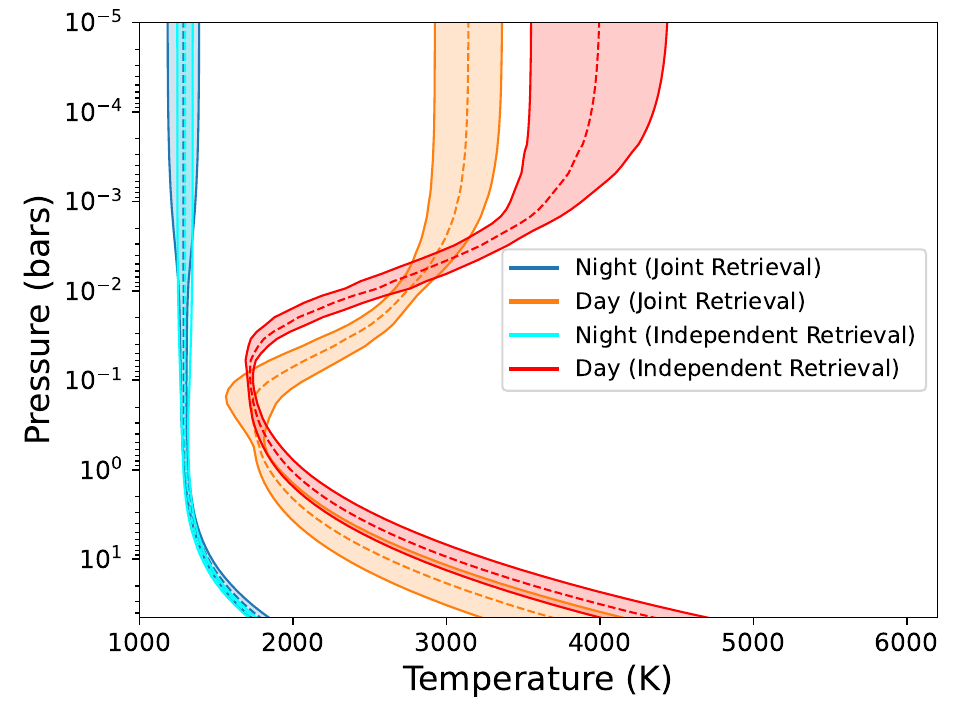}

\caption{Results of the joint fit to WD-0137B. All phases are fit simultaneously with a day- and night-TP profile combined according to Equation~\ref{eq:etadelta}. The left figure shows the retrieved best-fit spectra (with the dawn and pre-dawn phases not shown for clarity), while the right figure shows the dayside and nightside temperature-structure constraints compared to those from the independent, individual-phase retrievals.
\label{fig:wd0137_JOINT}}
\end{figure*}

Fits to four of the six phases with the joint phase curve retrieval are shown in Figure~\ref{fig:wd0137_JOINT}, alongside the temperature profiles of the day and nightside spectral components. Dayside and nightside abundances also agree with the independent retrievals to within 1-$\sigma$. The agreement between the joint and independent retrievals suggests that fitting the phases jointly is a justifiable approach for datasets such as this.

\subsection{EPIC 2122B}

\begin{figure*}[ht]
\plotone{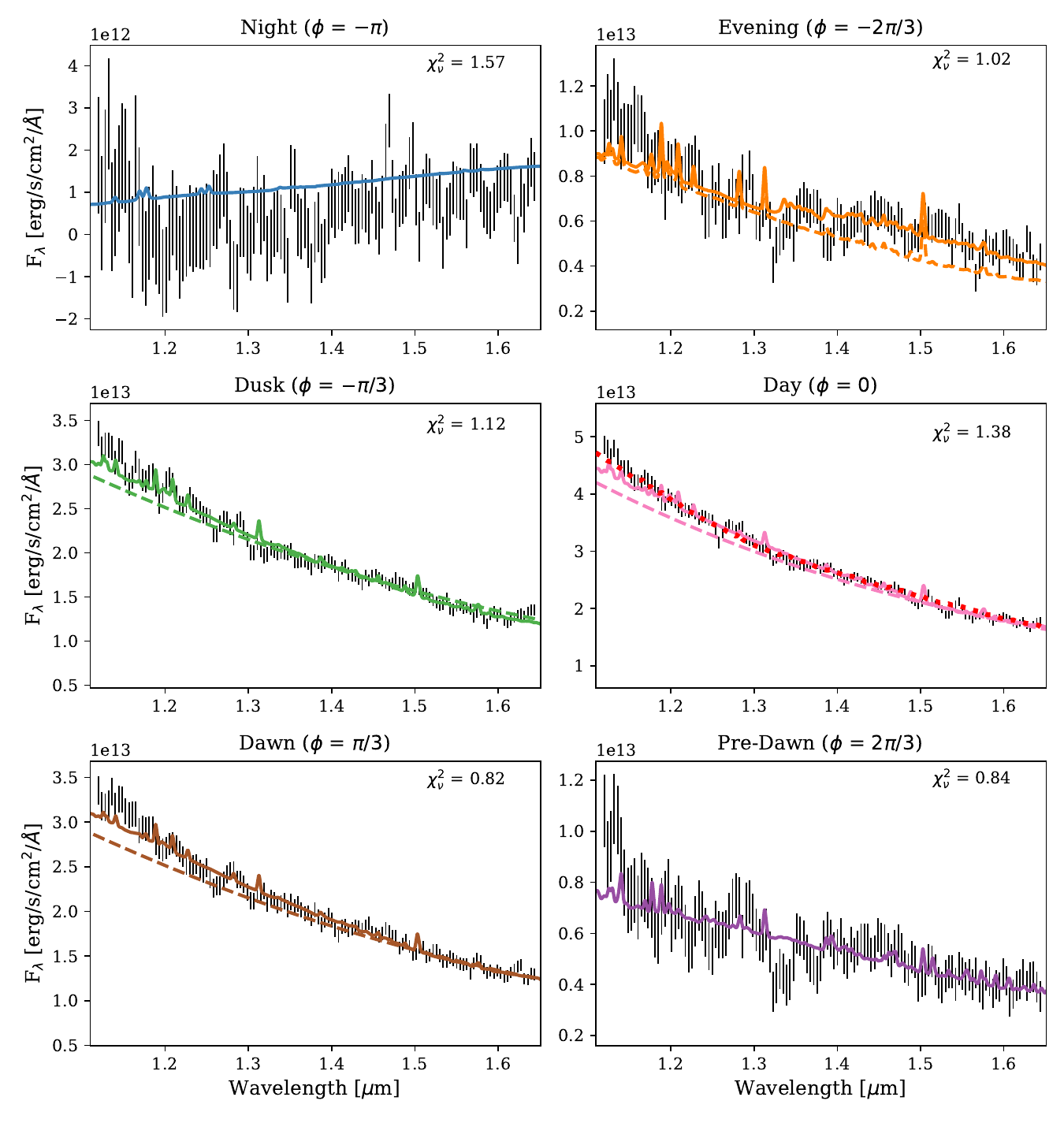}
\caption{Best-fit spectra of EPIC-2122 from fiducial retrievals to each of the six hemispheres. The evening phase, the only phase at which H$_2$O was detected, includes a fit without H$_2$O to demonstrate the spectral feature being fit. The fit to the dayside phase includes the best-fit from the retrieval with a dilution factor (dotted red line), which was statistically justified (see text). The dawn, dusk, and dayside phases also include, respectively, a 4,000, 4,000, and 4,500~K blackbody (dashed line) for reference.
\label{fig:EPIC2122_fits}}
\end{figure*}

Fiducial fits (a single, independent 5-parameter TP-profile, with uniform vertical abundances) to the six phase-resolved spectra of EPIC-2122B are shown in Figure~\ref{fig:EPIC2122_fits}. As with WD-0137B, the $\chi^2_\nu$ of the EPIC-2122B best fits indicate the spectra are well-fit, ranging from $\chi^2_\nu=0.82$ to 1.57. Again, because some phases' $\chi^2_\nu$ falls below one, the observational uncertainties appear to be overestimated, at least for those phases.

\begin{figure}[ht]
\epsscale{1.15}
\plotone{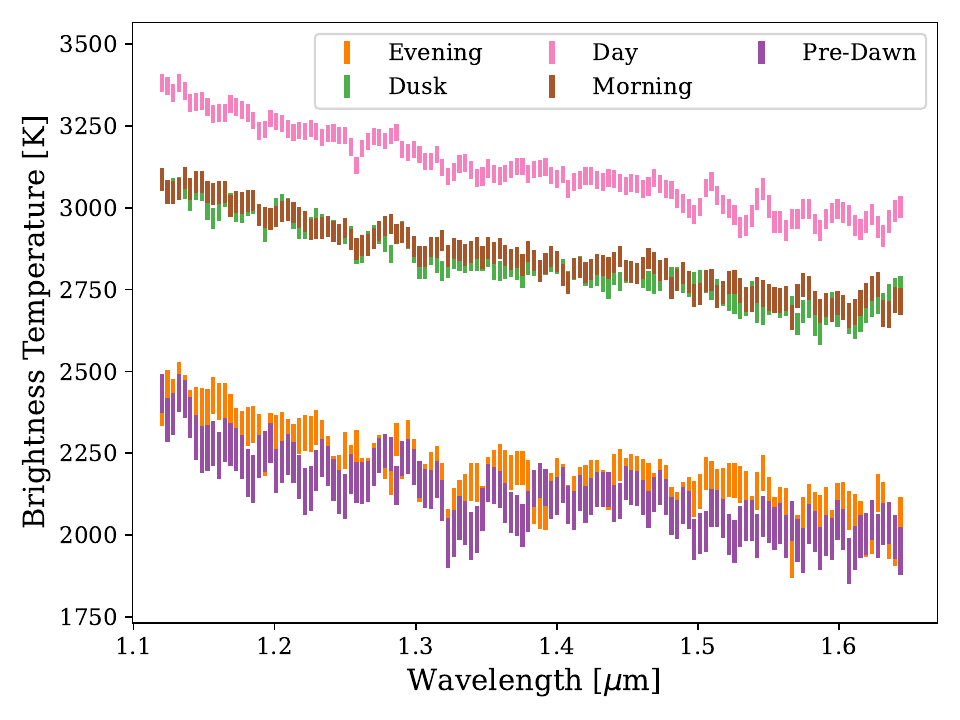}
\caption{Observed brightness temperatures of EPIC-2122B as a function of wavelength for each of the six hemispheres. Even though the spectra look featureless, the brightness temperatures imply a non-isothermal atmosphere. With H$^-$'s opacity, this indicates an inverted atmosphere. Nightside is not shown due to the number of wavelength bins with non-detections of the brown dwarf flux.
\label{fig:EPIC2122_BTs}}
\end{figure}

As was first described in \cite{zhou:2022}, the spectra of EPIC-2122B are devoid of apparent molecular absorption, most obviously from H$_2$O. This does not come as a surprise since thermal dissociation of molecules occurs at the very high temperatures associated with this object (T$_{irr} \approx 3,450$). Nonetheless, the spectrum of EPIC-2122B is not just a blackbody and exhibits a spectral slope consistent with H$^-$ opacity in an inverted atmosphere, similar to the dayside of WD-0137B. This is shown in the brightness temperatures of the spectra at each phase in Figure~\ref{fig:EPIC2122_BTs}, which have brightness temperatures that are hundreds of Kelvin hotter around 1.1$\mu{m}$ than at 1.6$\mu{m}$. 

\begin{table*}
\centering
\caption{EPIC-2122B retrieval scenario comparison.\label{tab:epic:scenarios}}
\begin{tabular}{ c||c|c|c|c|c}
\hline
 \hline
 Scenario & N$_{params}$ & -log($\mathcal{L}$) & $\Delta$BIC & $\Delta$AIC & Conclusion \\
 \hline
 Fiducial            & 7   & -3747.3  & --   &   --   & --                            \\
 Non-uniform e$^-$   & 9   & -3747.3  & -9.7 & -4.0   & Not Justified                 \\
 Non-uniform H$_2$O  & 9   & -3747.3  & -9.7 & -4.0   & Not Justified                 \\
 Dual-TP             & 15  & -3720.4  & 24.7 & 41.8   & Strongly Justified           \\
 Fill Factor         & 8   & -3714.2  & 61.3 & 64.2   & Most strongly justified \\
 \hline
\end{tabular}
\end{table*}

\subsubsection{Temperature Structures}

As described above, the fact that the brightness temperatures decrease from short to long wavelength between 1.1 and 1.6$\mu{m}$ is indicative of a temperature inversion in the atmosphere of EPIC-2122B. This is because H$^-$ opacity, which will dominate the continuum at these high temperatures \citep[e.g.,][]{arcangeli:2018}, also decreases from  1.1 to 1.6$\mu{m}$, which implies that temperatures where the contribution function is higher in the atmosphere (i.e., at lower pressure) are hotter. Interestingly, the brightness temperatures shown in Figure~\ref{fig:EPIC2122_BTs} decrease from 1.1 and 1.6$\mu{m}$ at all phases except night ($\phi = \pi$), where the spectrum of the brown dwarf companion is not well-detected. In Section~\ref{sec:GCMs}, we discuss explanations for how the nightside atmosphere can appear to remain inverted. 

\begin{figure}[ht]
\epsscale{1.15}
\plotone{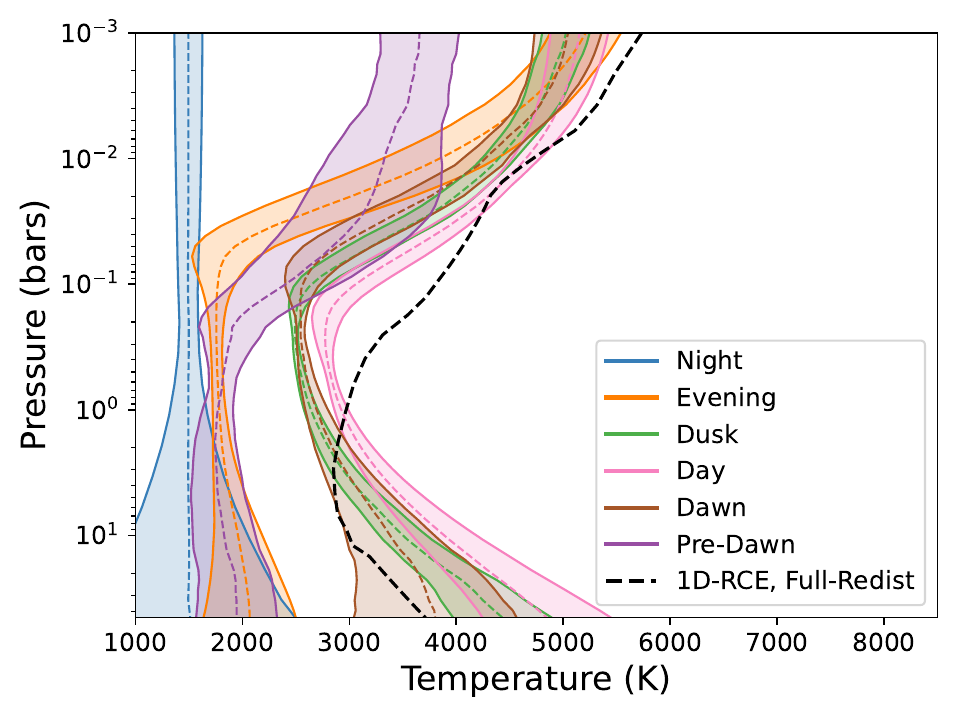}
\caption{Retrieved temperature structure constraints for EPIC-2122B for each of the six hemispheres compared to the temperature structure from a full-heat redistribution 1D radiative-convective equilibrium model.
\label{fig:EPIC2122_TPs}}
\end{figure}

This qualitative picture of the longitudinal temperature structure of EPIC-2122B is confirmed with the fiducial retrievals. Figure~\ref{fig:EPIC2122_TPs} shows the retrieved TP profiles at the 6 independently-fit phases. While cooler temperatures are seen before dawn and after dusk, inversions are still clearly seen in the atmosphere. The dayside-hemisphere retrievals roughly match the theoretical expectations from the full-heat redistribution 1D model from \cite{lothringer:2020c}, though the retrieved temperature profile{ (i.e., the photosphere and temperature inversion)} appear uniformly shifted towards lower pressures, as was similarly seen in WD-0137B.

\subsubsection{Abundances}

\begin{figure*}[ht]
\epsscale{1.15}
\plottwo{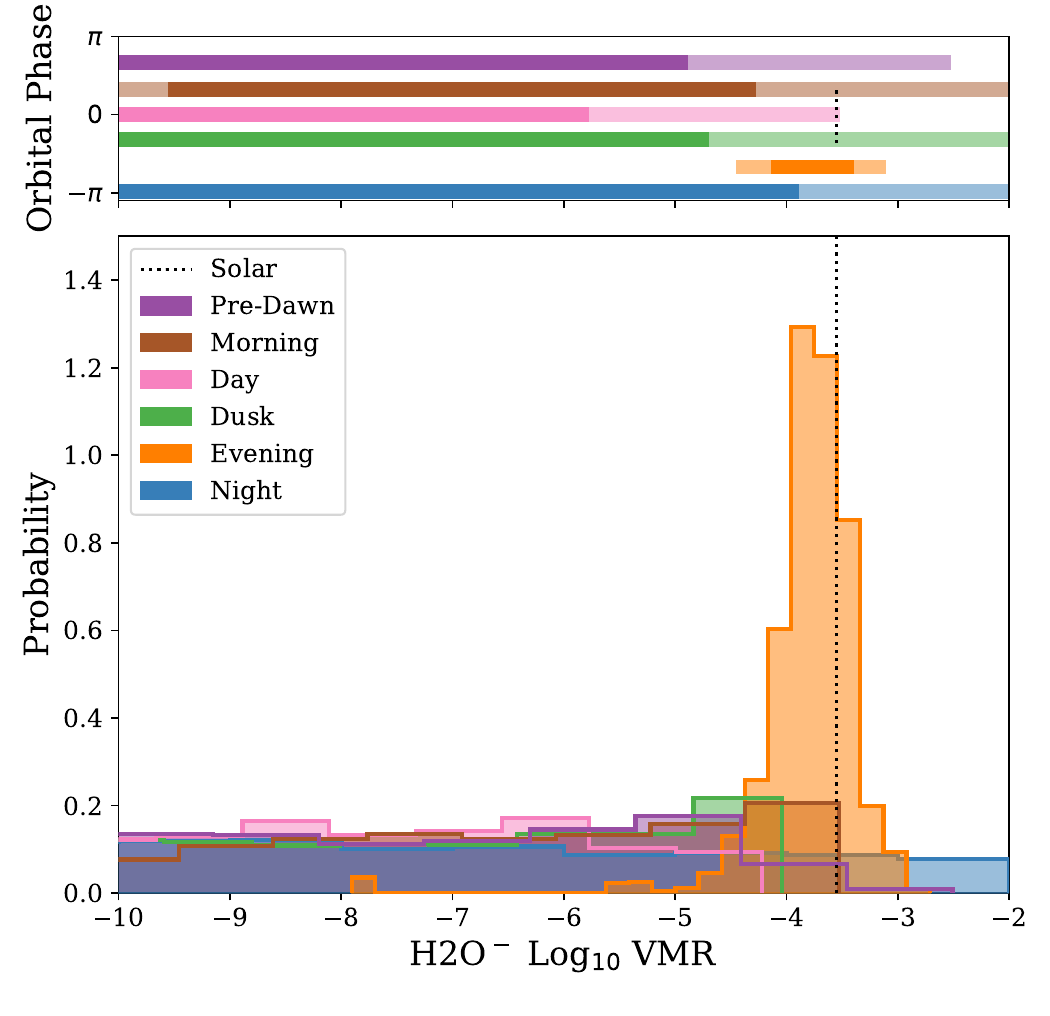}{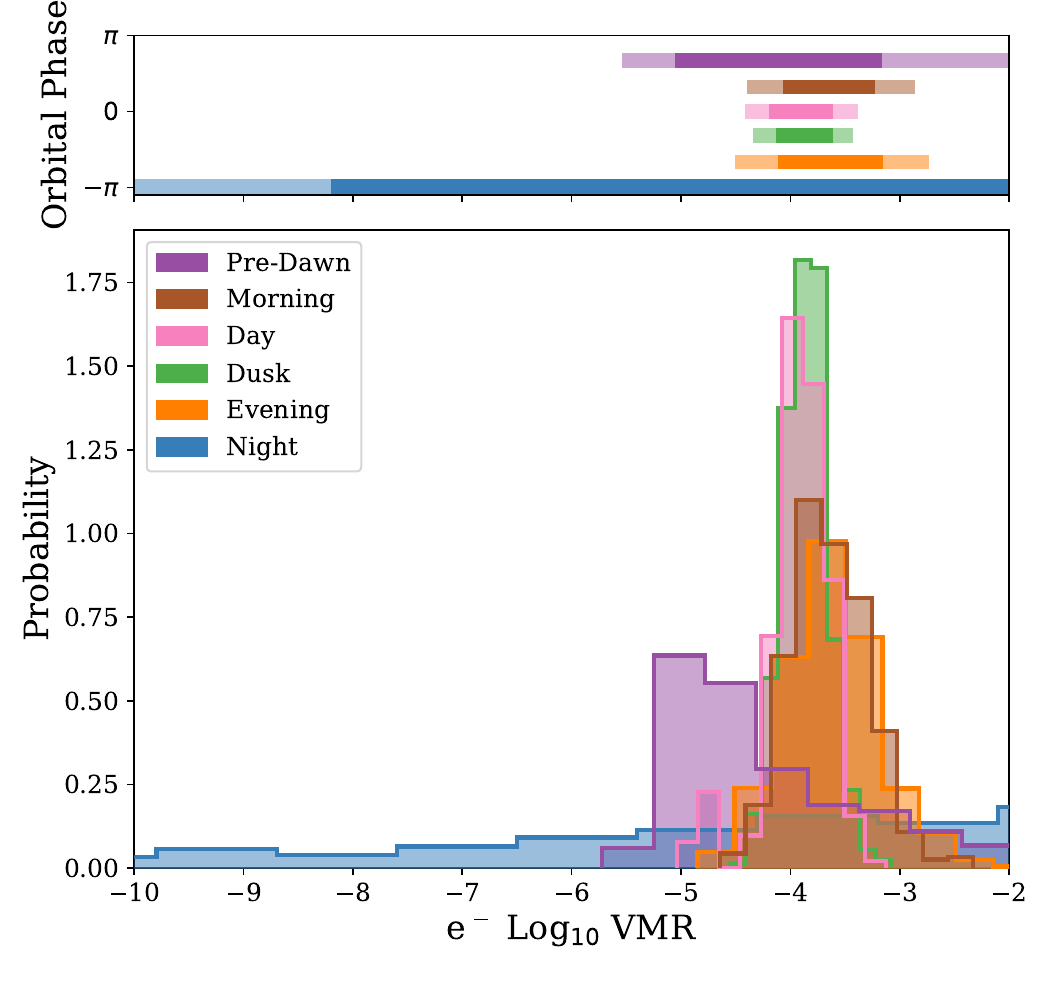}
\caption{Abundance of H$_2$O (left) and e$^-$ (right) in EPIC-2122B  for each of the six hemispheres. H$_2$O is poorly constrained for all phases because of thermal dissociation and the dominance of H$^-$ opacity except the evening phase ($\phi=\frac{1}{3}\pi$).
\label{fig:epic2122_abunds}}
\end{figure*}

Figure~\ref{fig:epic2122_abunds} shows the retrieved abundances of H$_2$O and e$^-$. H$_2$O is unconstrained at all phases except evening because of thermal dissociation and the dominance of H$^-$ opacity. The best-fit spectrum but without H$_2$O opacity is also shown in Figure~\ref{fig:EPIC2122_fits} for the evening phase, demonstrating that the retrieval does see a significant spectral feature. Curiously, this is not replicated at the corresponding pre-dawn phase. We note, however, that high H$_2$O abundances are allowed by the retrieval at nearly all phases, with only the dayside providing an 1-$\sigma$ solar metallicity upper-limit.

The e$^-$ abundance itself is retrieved to be present at all phases, except the poorly constrained anti-stellar night phase. This suggests that H$^-$ survives past the terminators and onto the nightside hemisphere. While the recombination of atomic H into H$_2$ has been shown to transport heat from day to night on similarly irradiated objects \citep{bell:2018,tad:2018,tan:2019b,mansfield:2019b}, these results suggest enough atomic H and free e$^-$ remains visible on the nightside hemisphere of EPIC-2122B for H$^-$ to continue as the dominant opacity source.


\subsubsection{Non-Fiducial Scenarios}

As with WD-0137B, we performed additional retrievals to investigate scenarios including non-uniform vertical abundances and non-homogenous spectra {(see Table~\ref{tab:epic:scenarios})}. Retrievals including non-uniform vertical abundances of H$_2$O were not justified, as was clear by the non-detection of H$_2$O in the dayside spectrum. A similar story holds for H$^-$ despite the clear spectral signatures of H$^-$, suggesting that the abundance of H$^-$ does not vary much throughout the pressure range probed by these observations.

Unlike for WD-0137B, retrievals including both a dual-temperature structure scenario and a filling/dilution factor are favored by the data. \cite{zhou:2022} found a similar need for a filling factor when comparing 1D self-consistent models to the dayside spectrum. The dual-TP scenario resulted in an increase in the log-likelihood of 26.9 and a $\Delta$BIC of 24.7, indicating the additional parameters necessary to model the non-homoegenity were justified by the data. {The dual-TP scenario was poorly converged due to a very low acceptance rate of samples, however the fit for this scenario was improved enough compared to the fiducial retrieval to be considered strongly justified with a $\Delta$BIC of 24.7.} The filling-factor scenario increased the log-likelihood by a further 6.2, resulting in a $\Delta$BIC of 61.3. Therefore, the scenario preferred by the data includes a filling/dilution factor treatment.

The fact that the dayside filling factor scenario was able to find such a good fit compared to the fiducial (see Figure~\ref{fig:EPIC2122_fits}) and dual-TP scenarios indicates that the dayside spectrum of EPIC-2122B is dominated by a hotspot, as has been suggested by GCMs, observations, and retrievals of hot Jupiters \citep{showman:2009,majeau:2012,parmentier:2018,arcangeli:2019,taylor:2020,beltz:2021}. The hotter temperatures this allowed in turn increased H$^-$ opacity and steepened the blue-ward slope of the retrieved spectra to better fit the observations. This also suggests that the contribution from the nightside hemisphere due to EPIC-2122B's orbital inclination did not significantly affect the dayside atmosphere because this would have induced a lower overall temperature (though other phases may have been more affected, see Section~\ref{sec:GCMs}). By including the filling factor, which was found to equal 0.17 $\pm$ 0.013, the retrieval could increase the temperature of the planet until $\beta$ = 1.52 $\pm$ 0.25 (i.e., the mean atmospheric temperature from irradiation was $\sim$50\% higher than the full-heat redistribution equilibrium temperature) resulting in maximum temperatures at the photosphere of about 6,000~K.


\subsubsection{Joint Phase Curve Fit}

\begin{figure*}[ht]
\epsscale{1.15}
\plottwo{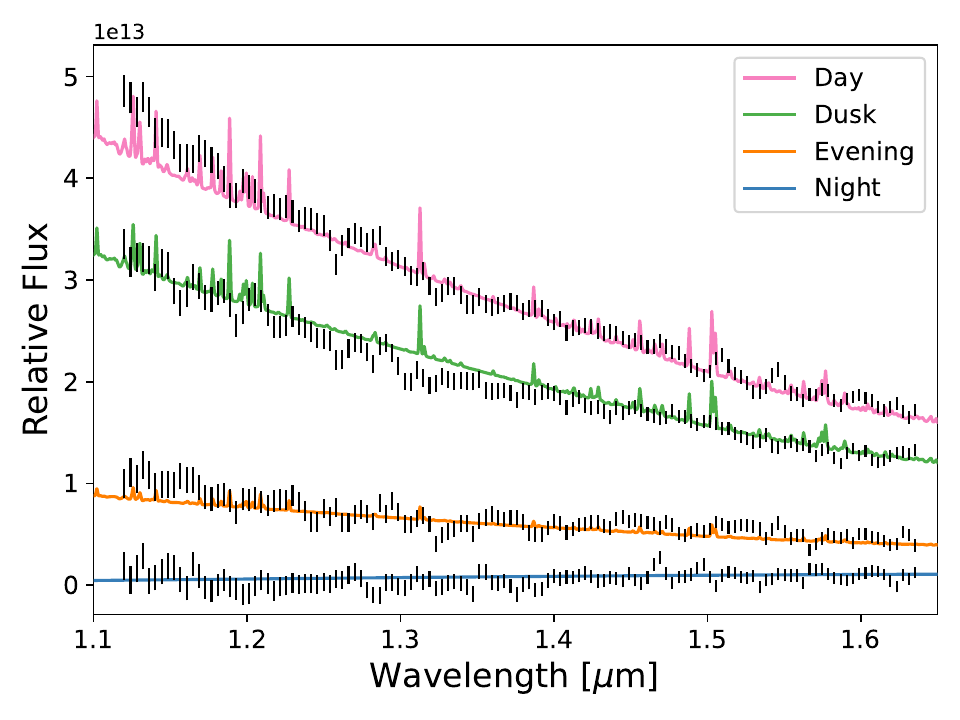}{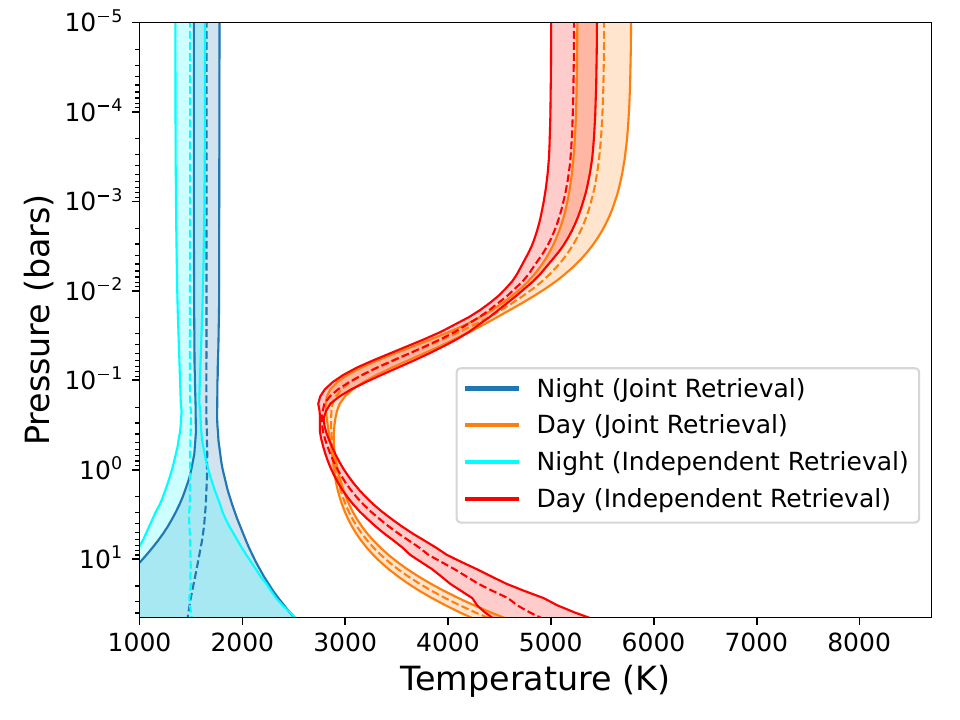}

\caption{Results of the joint fit to EPIC-2122B. All phases are fit simultaneously with a day- and night-TP profile combined according to Equation~\ref{eq:etadelta}. The left figure shows the retrieved best-fit spectra, while the right figure shows the dayside and nightside temperature-structure constraints compared to those from the independent, individual-phase retrievals.
\label{fig:epic2122_JOINT}}
\end{figure*}

We also carried out a joint retrieval of the full phase curve, as was done with WD-0137B. With our simple 2-component fit (a dayside and nightside hemisphere), we can match the observations adequately, though the short-wavelength portion of the dayside spectrum is not well fit (see Figure~\ref{fig:epic2122_JOINT}). The retrieved phase curve offset is consistent with zero at $c = 0.0016 \pm 0.003$, while the day-night gradient is represented by $\delta = 1.22 \pm 0.016$ (see Equation~\ref{eq:etadelta}), suggesting a similarly symmetric and steeper-than-sinusoid phase curve as WD-0137B (see Section~\ref{sec:wd0137:pcfit}). Again, this is consistent with the Fourier analysis of \cite{zhou:2022}.

\section{Discussion}

\subsection{Comparison to GCMs}\label{sec:GCMs}

To understand how our 1D retrievals of the phase-resolved spectra compare to 3D predictions, we compared our fiducial retrieved temperature structures to the global circulation models (GCM) used in \cite{zhou:2022}, which were in turn based on the models of \cite{tan:2019b}. These idealized semi-gray, two-stream GCM use a two-channel irradiation scheme specific to these irradiated brown dwarfs to account for the large incoming UV flux from the hot WD primary companions. \cite{zhou:2022} found that the shapes of the phase curves for both WD-0137B and EPIC-2122B were well-matched by these GCMs. See the appendix of \cite{zhou:2022} for more details on the GCMs used here.

\begin{figure*}[ht!]
\epsscale{1.0}
\plotone{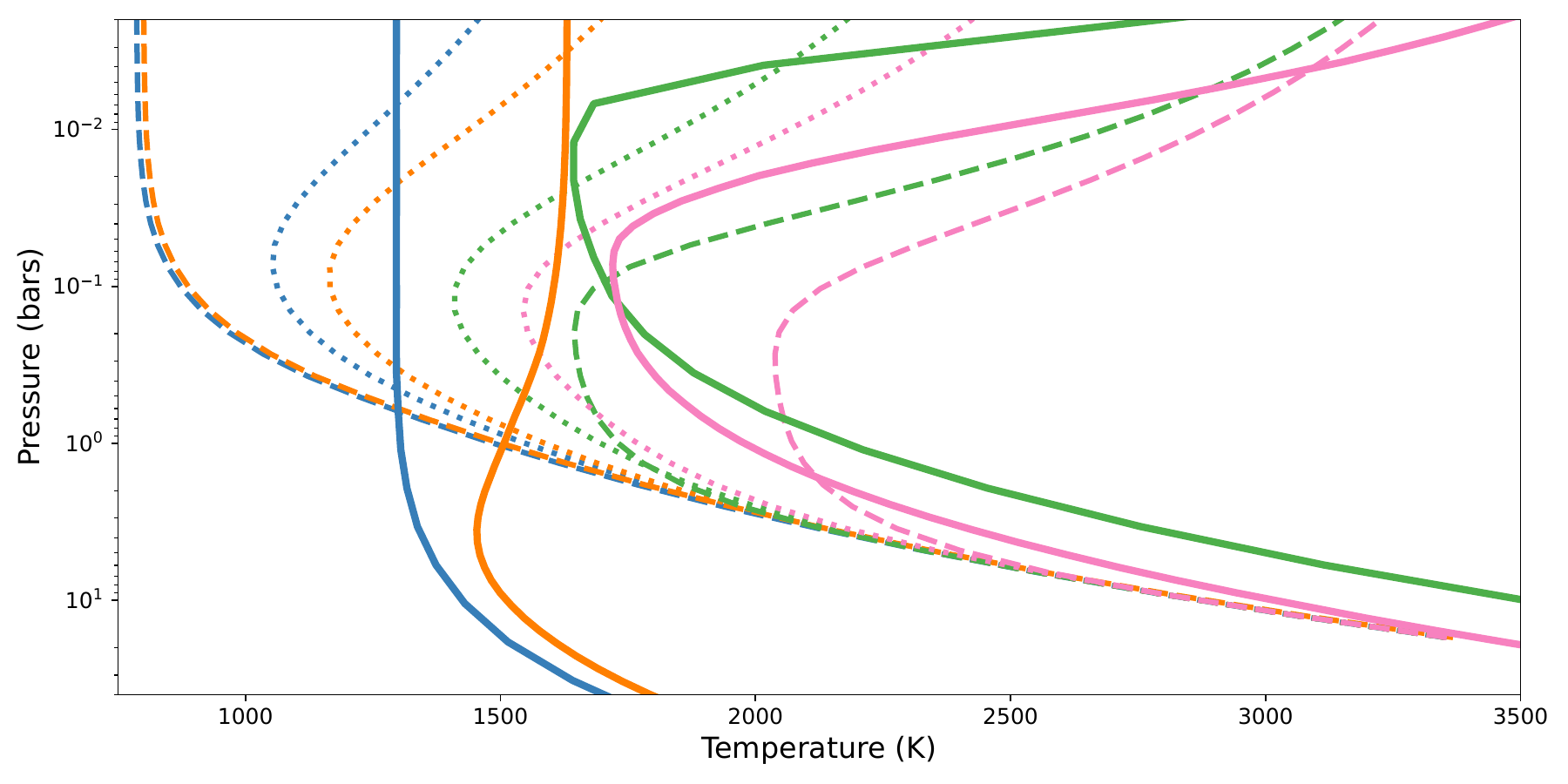}\plotone{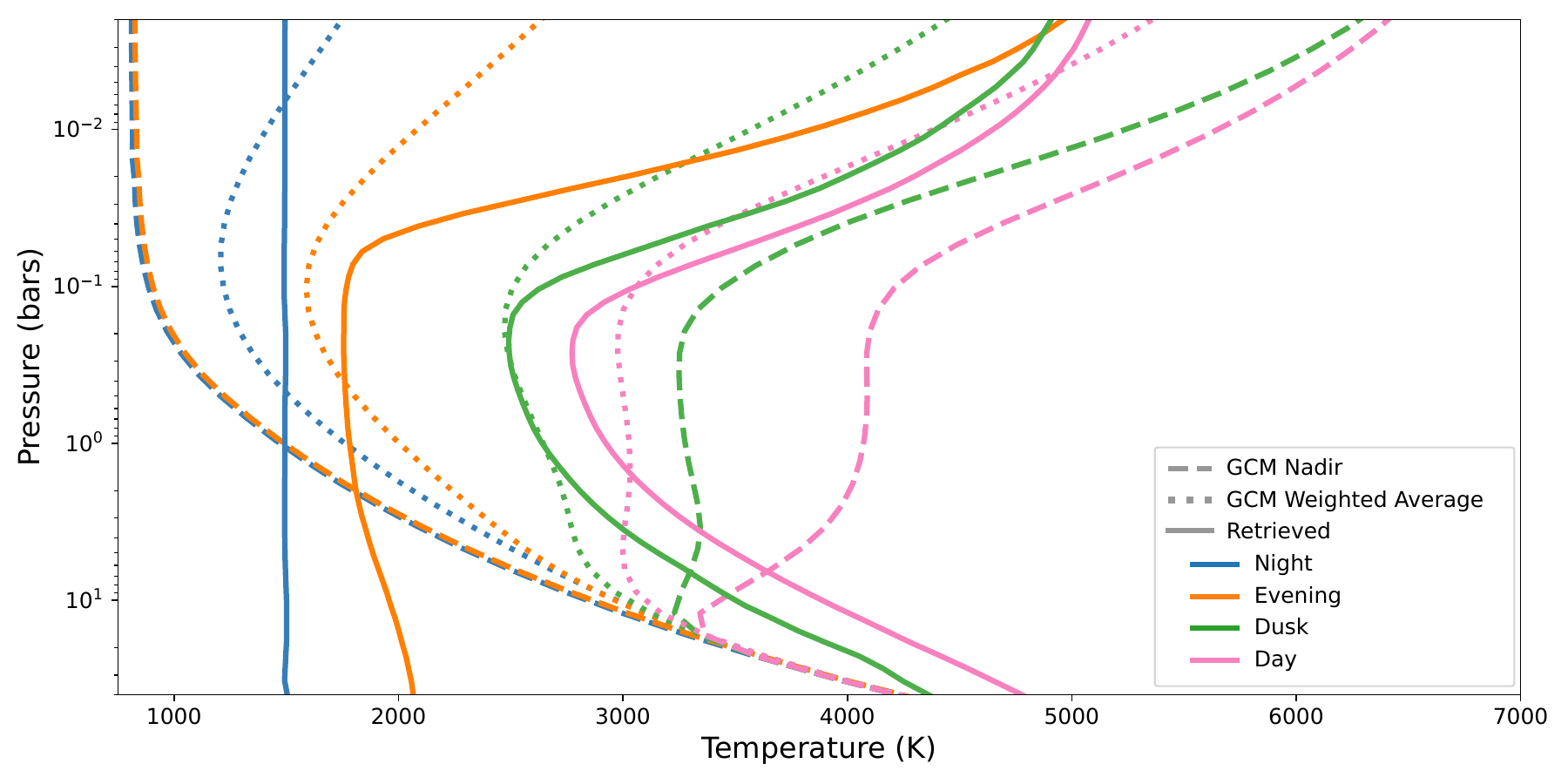}
\caption{Temperature structures from the global circulation model from \citep{zhou:2022} for WD-0137B (top) and EPIC-2122B (bottom). Dashed lines are structures from the point on the brown dwarf directly nadir to observer, while dotted lines are a projection weighted average of all visible points on the brown dwarf. Solid lines are the best-fit temperature structures from the fiducial retrievals. Dawn and pre-dawn phases are omitted for clarity, but closely follow the structures for the dusk and evening phases, respectively, in both the retrieval and GCM. 
The solid black line also indicates the temperature structure of a self-consistent 1D atmosphere model of the dayside atmosphere from \citep{lothringer:2020c}, with the dotted black line indicating the normalized contribution function in the WFC3/G141 bandpass for that model.  
\label{fig:GCM:TP}}
\end{figure*}

Figure~\ref{fig:GCM:TP} compares the temperature structures from the GCMs (dashed and dotted) to the retrieved and 1-D modeled TP profiles (solid lines). We first compare the temperature structure of the nadir point on the brown dwarf (i.e., the location on the brown dwarf directly pointed towards Earth), given as dashed lines. At WD-0137B and EPIC2122B's inclinations of 35$^{\circ}$ \citep{maxted:2006} and 56$^{\circ}$ \citep{casewell:2018}, respectively, this will correspond to a latitude of 55$^{\circ}$ for WD-0137B and 34$^{\circ}$  for EPIC-2122B. The deep atmosphere temperatures do show significant deviations between the retrieval and GCM, but this is likely be due to a combination of 1) the fact that the WFC3 observations do not constrain the temperature structure at pressure above 10 bars and 2) the GCMs assumed fixed effective temperatures of 3,500 and 4,500~K for WD-0137B and EPIC-2122B, respectively. In future studies, the retrieved temperature structures can be used as inputs to GCMs for more one-to-one comparisons.

In general, the nadir GCM temperature structures are too warm on the dayside and too cool on the nightside. This is because the rest of each hemisphere is contributing to the observed and retrieved temperature structure{, effectively ``muting" the temperature extremes}. For example, the nadir point will be at the same longitude as the sub-stellar point for the dayside phase; however, the observed flux will be diluted from locations off the sub-stellar longitude and we should expect the retrieved temperature structure to be somewhat cooler than the nadir dayside GCM temperature structure and this is indeed what is seen.{ Similarly, the nightside will see contribution from the high-latitudes of the dayside, so we would expect the observed nightside temperatures to be warmer than the nadir nightside GCM profile.} This idea has been explored in the context of GCMs of non-transiting planets like Ups And b \citep{malsky:2021}.

To account for the contribution of the whole visible hemisphere at any given phase, we also show in Figure~\ref{fig:GCM:TP} the weighted average of the temperature profiles from all visible points on the brown dwarf, given as dotted lines. The weights are determined by the cosine of the central angle between the nadir point (which will have the highest weight) and the considered location from the GCM. These weighted average temperature profiles will be cooler than the nadir profiles on the dayside hemisphere and warmer than the nadir profiles on the nightside. 

In the end, our retrieved temperature profiles indeed lie somewhere between the nadir and weighted-average temperature profiles. These weighted-average profiles help us understand why the nightside-hemisphere phases (i.e., evening and pre-dawn) showed temperature inversions in both WD-0137B and EPIC-2122B. While the nadir location at the evening and pre-dawn phases is non-irradiated, there is still flux being contributed from locations on the dayside, irradiated hemisphere. This means that there was always contribution from the high-latitude dayside, which for EPIC-2122B was enough to drive the observed spectrum to show evidence of H$^{-}$ opacity and H$_2$O dissociation and the retrieved temperature profile to be inverted.

\begin{figure*}[ht]
\gridline{\fig{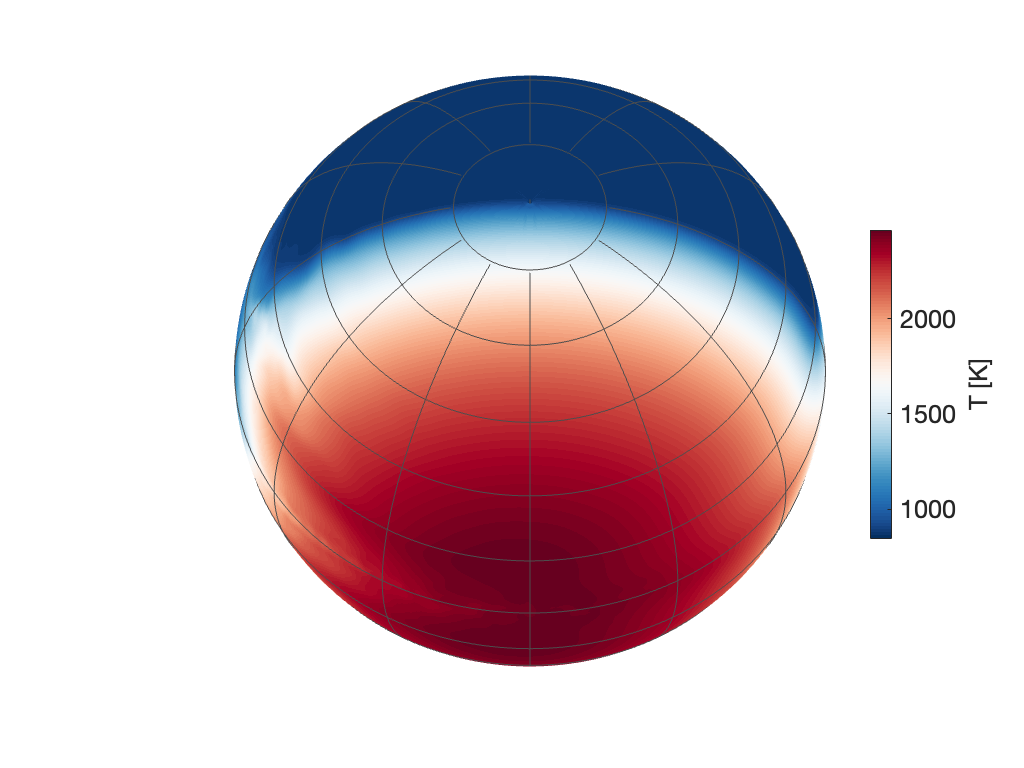}{0.45\textwidth}{(a) 1.0 bar - Dayside}
          \fig{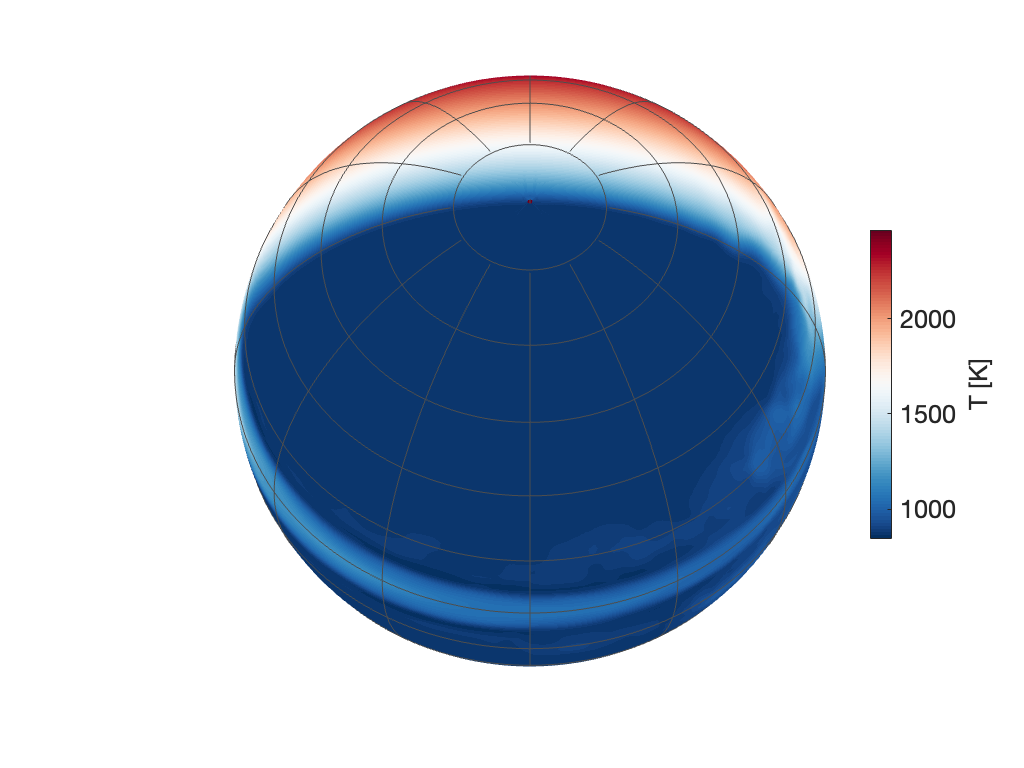}{0.45\textwidth}{(b) 1.0 bar - Nightside}}
          
\gridline{\fig{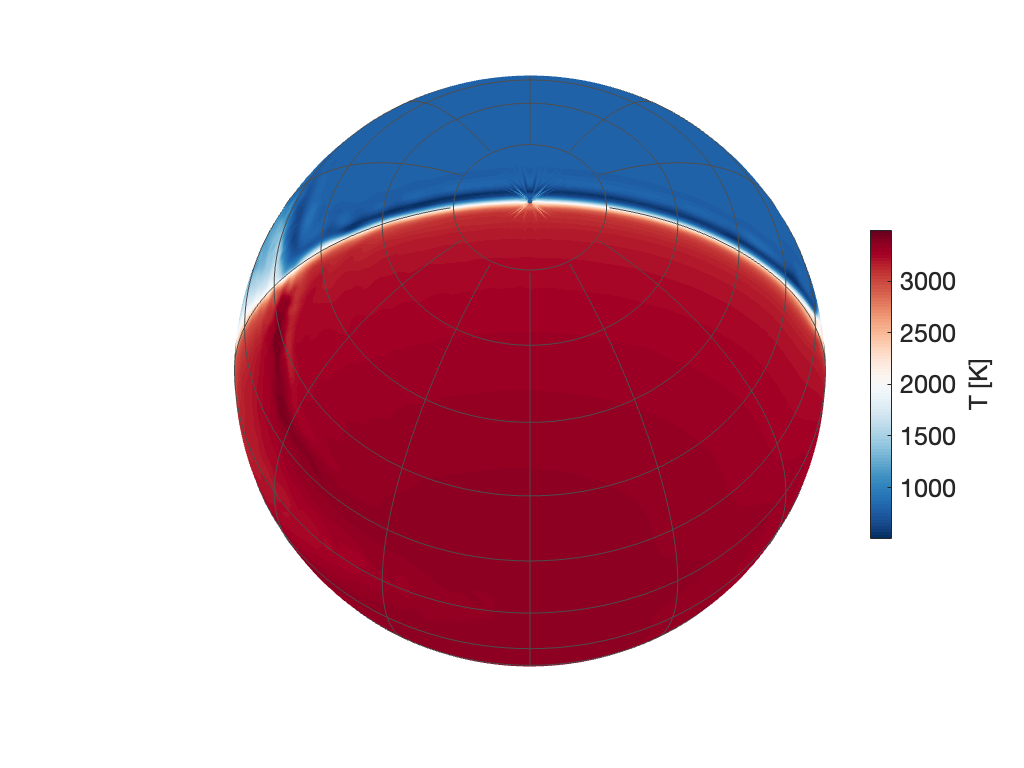}{0.45\textwidth}{(c) 0.01 bar - Dayside}
          \fig{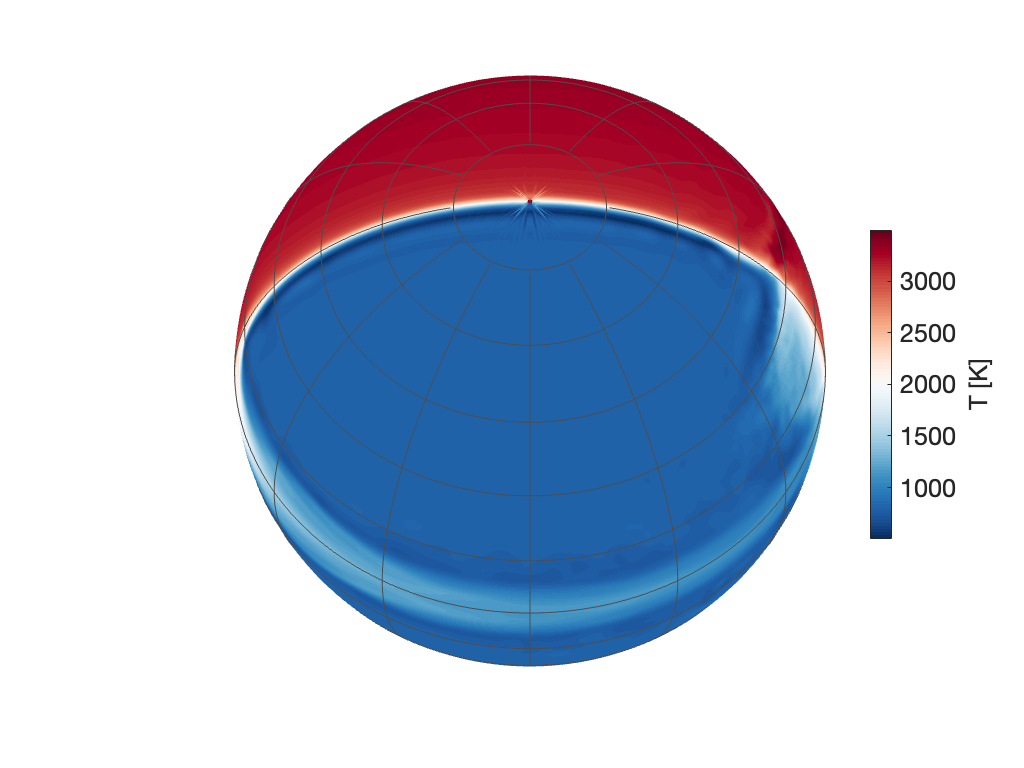}{0.45\textwidth}{(d) 0.01 bar - Nightside}}
          
\gridline{\fig{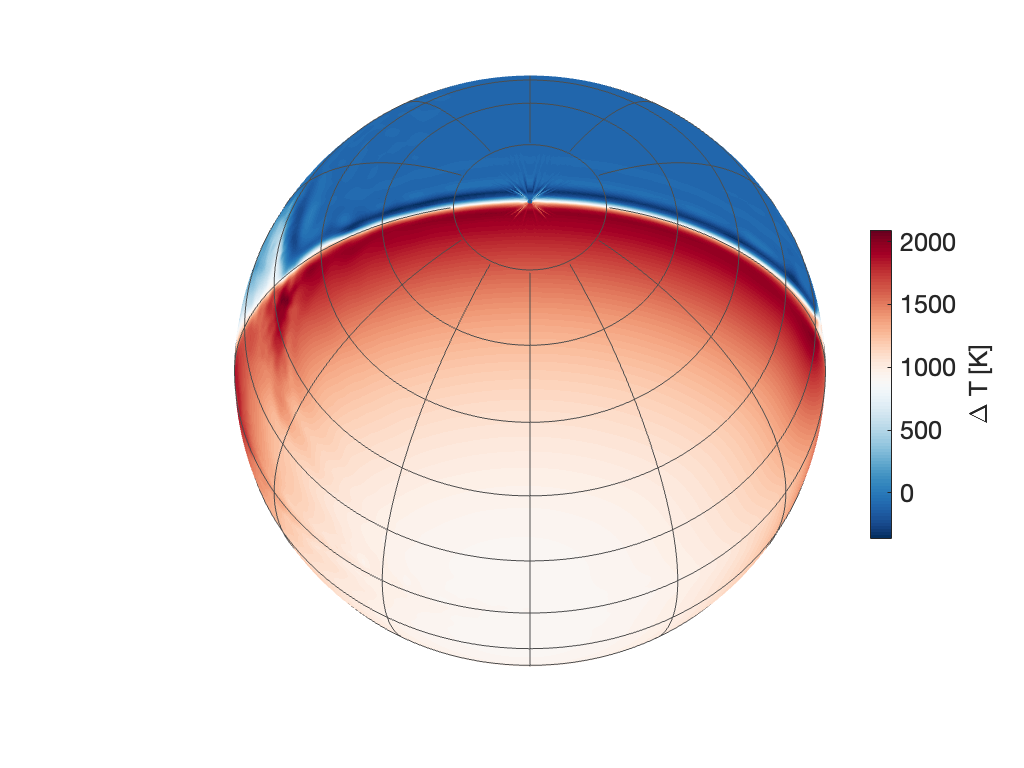}{0.45\textwidth}{(e) 0.01-1.0 bar Difference - Dayside}
          \fig{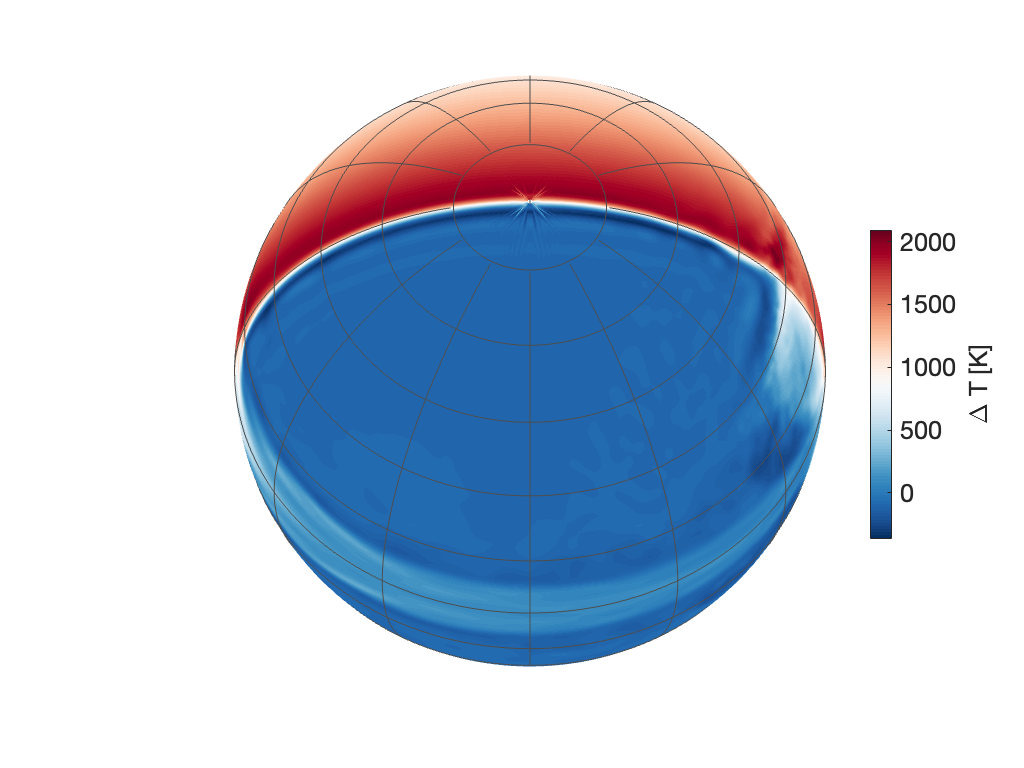}{0.45\textwidth}{(f)  0.01-1.0 bar Difference - Nightside}}
\caption{The temperature map of WD-0137B at 1 bar, 0.01 bar, and their difference on both the dayside and nightside hemispheres as viewed from Earth.
\label{fig:WD0137:GCMglobe}}
\end{figure*}

\begin{figure*}[ht]
\gridline{\fig{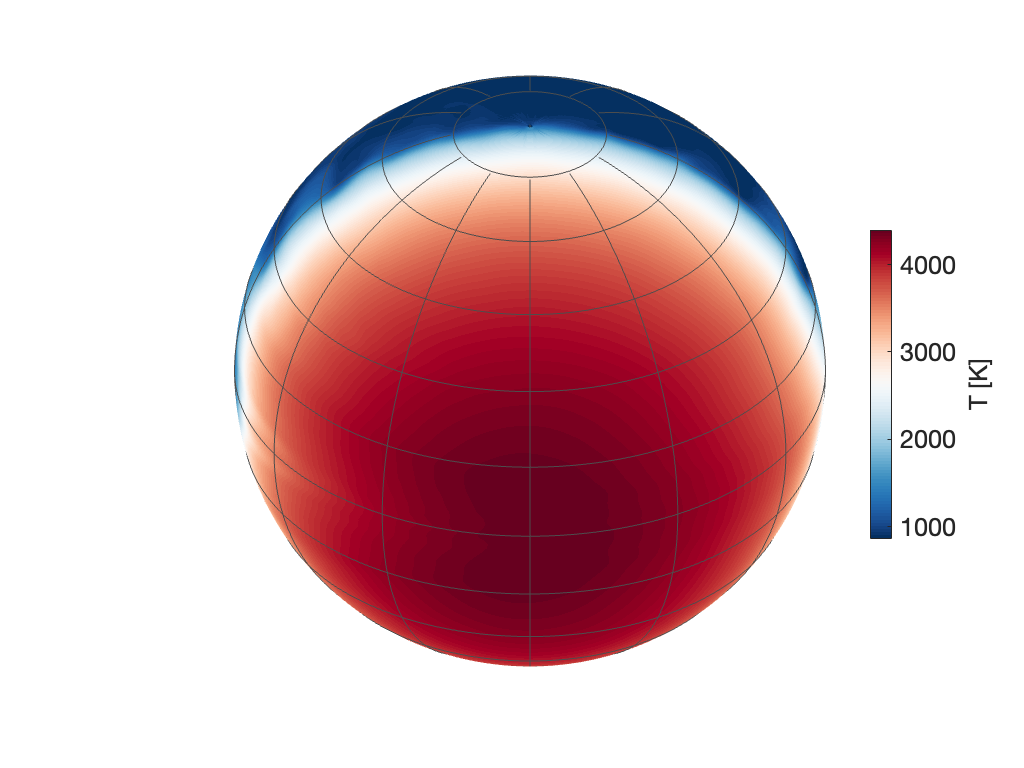}{0.45\textwidth}{(a) 1.0 bar - Dayside}
          \fig{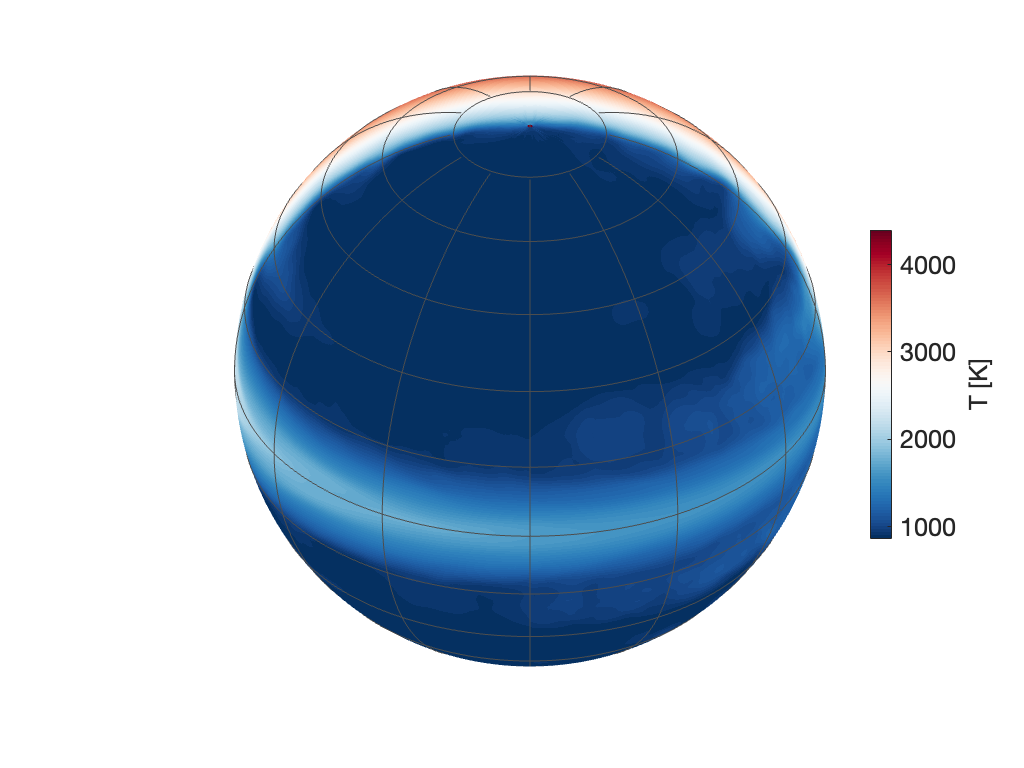}{0.45\textwidth}{(b) 1.0 bar - Nightside}}
          
\gridline{\fig{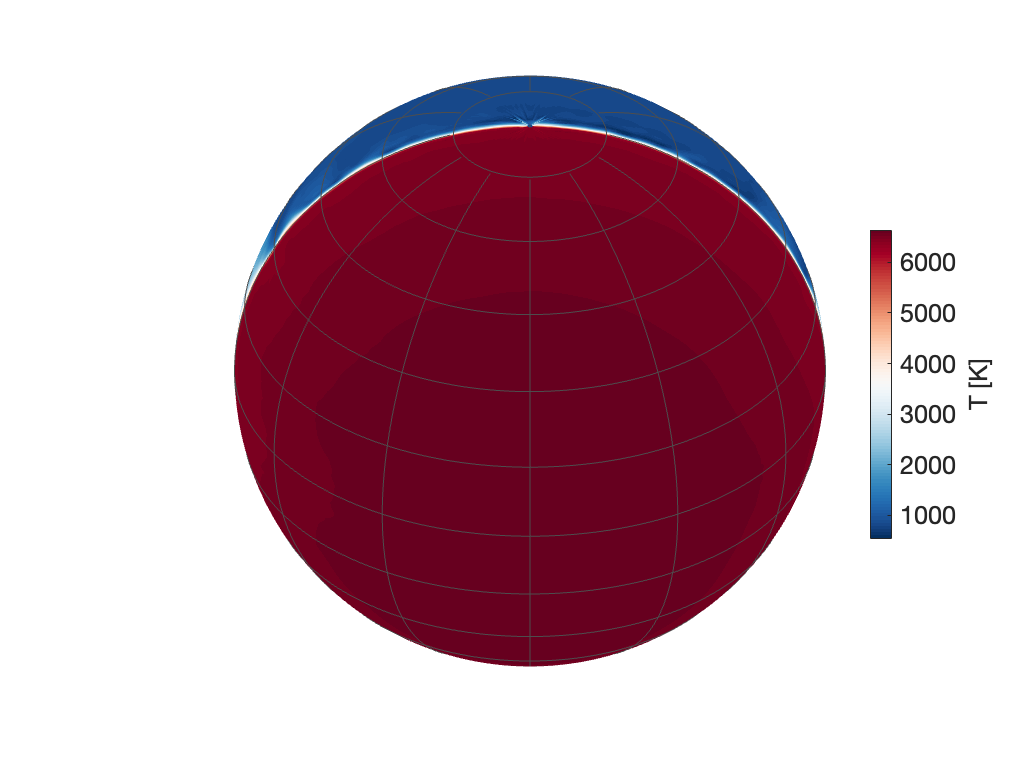}{0.45\textwidth}{(c) 0.01 bar - Dayside}
          \fig{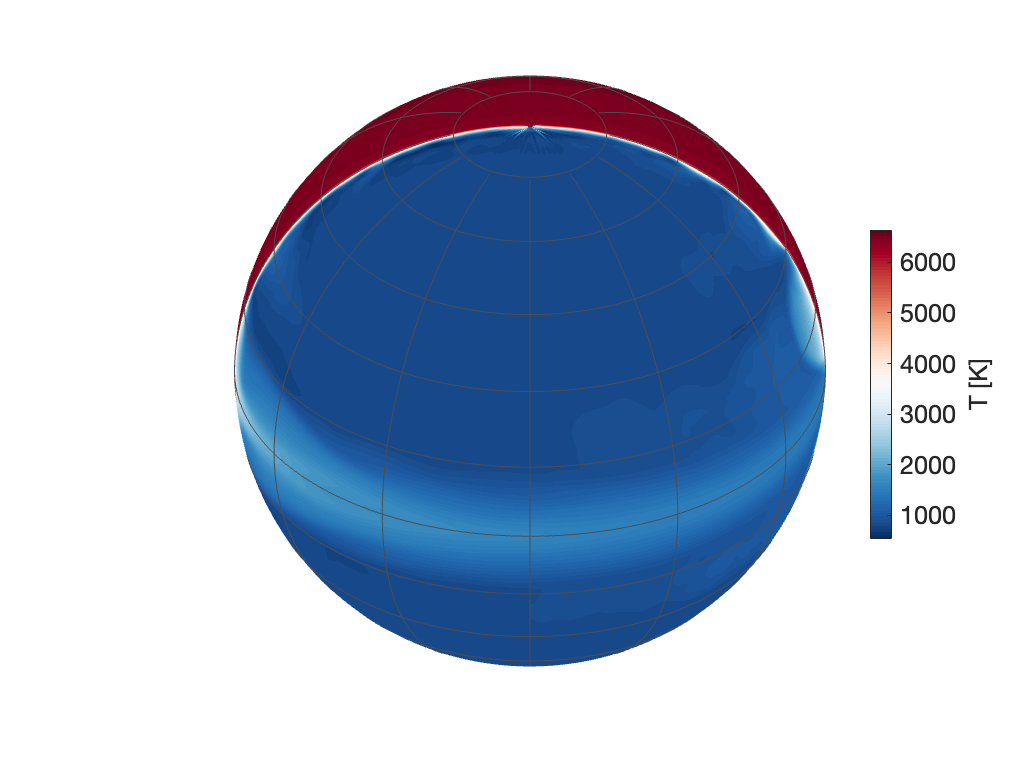}{0.45\textwidth}{(d) 0.01 bar - Nightside}}
          
\gridline{\fig{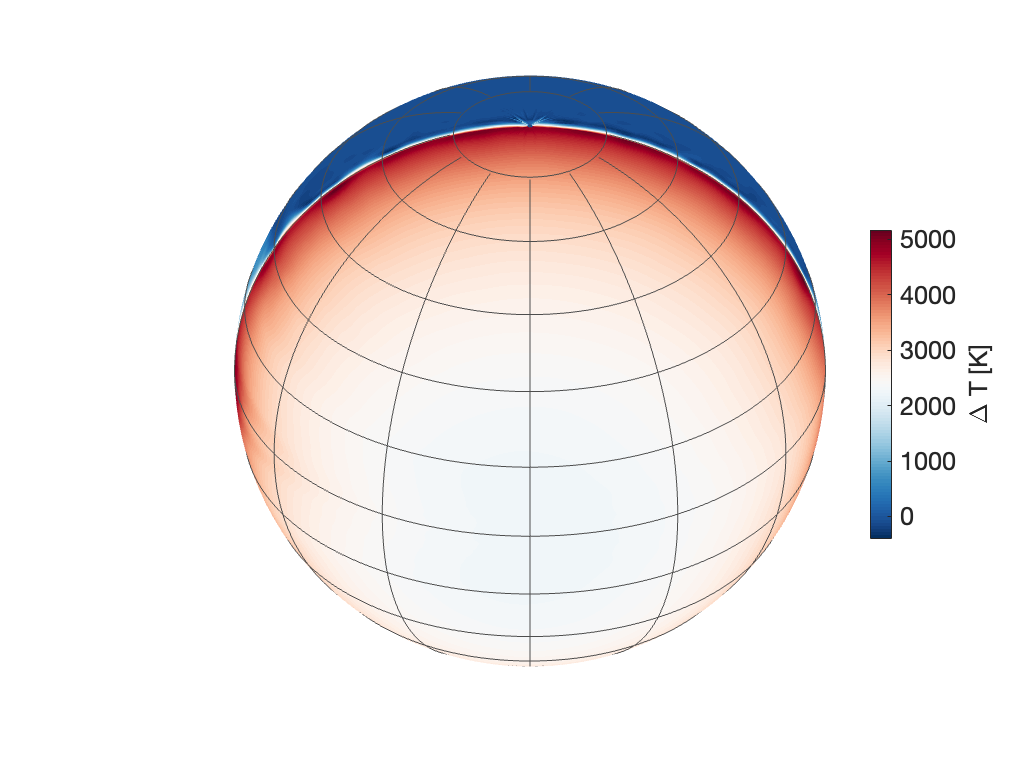}{0.45\textwidth}{(e) 0.01-1.0 bar Difference - Dayside}
          \fig{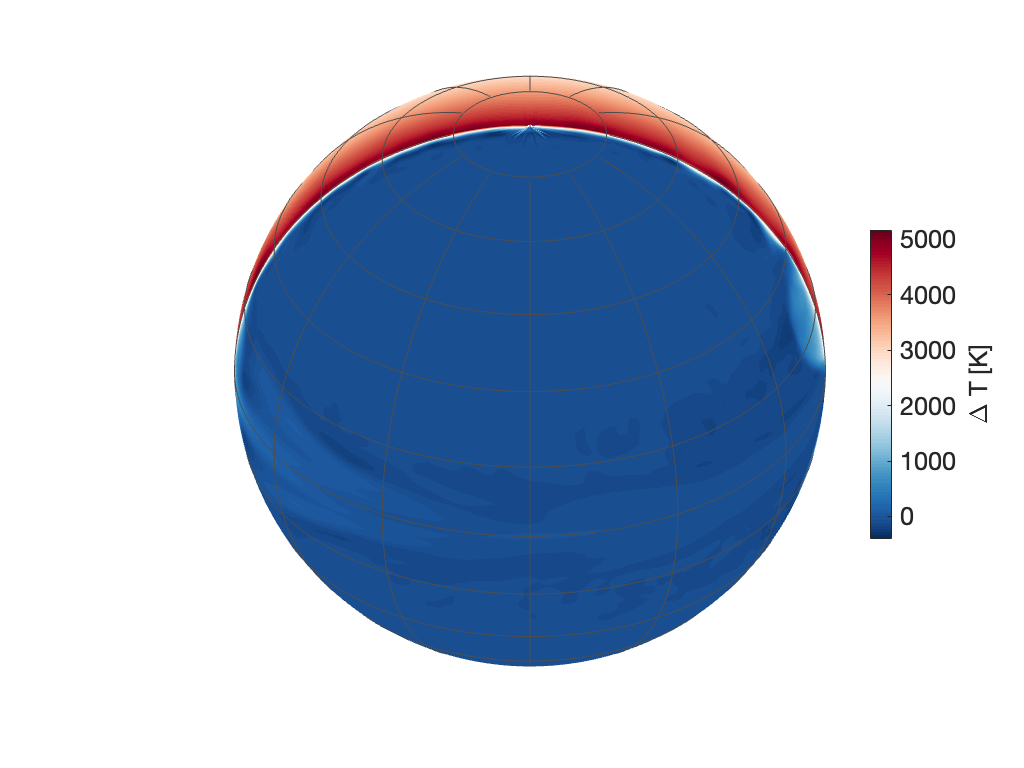}{0.45\textwidth}{(f)  0.01-1.0 bar Difference - Nightside}}
\caption{Same as Figure~\ref{fig:WD0137:GCMglobe}, but for EPIC-2122B.
\label{fig:EPIC2122:GCMglobe}}
\end{figure*}

{To further demonstrate the contribution from the whole visible hemisphere at a given phase, Figures~\ref{fig:WD0137:GCMglobe} and \ref{fig:EPIC2122:GCMglobe} show temperature maps of WD-0137B and EPIC-2122B from the GCM. Also shown is the temperature difference between 0.01 bar and 1 bar, where any positive value indicates a temperature inversion. As can be seen, the entire dayside hemisphere exhibits an inverted temperature profile. Importantly, due to the orbital inclination, a large portion of the polar regions of each object are always visible, no matter the phase. Furthermore, these polar regions are inverted for both objects with a 0.01 - 1.0 bar temperature difference of up to 2,000 and 5,000~K for WD-0137B and EPIC-2122B, respectively. This helps to explain why the retrieved temperature profiles for the evening and pre-dawn phases indicate significant temperature inversions, despite the nadir point being beyond the day-night terminator. The larger magnitude of the polar inversion for EPIC-2122B helps explain why the evening and pre-dawn TP profiles remain more strongly inverted than compared to WD-0137B.}

{While the inverted-nature of the retrieved temperature structures can be understood as a result of the orbital inclination, we can also qualitatively compare the overall shape. On the dayside hemisphere, EPIC-2122B's GCM predictions suggest a larger isothermal region between the internal adiabat and the temperature inversion, extending nearly 2 dex in pressure. Meanwhile, this ``tropopause" region is much smaller in the WD-0137B GCMs, extending only about a dex in pressure, driven by the proportionally higher contribution of WD-0137B's internal temperature to the total effective temperature. In all cases, the retrieved temperature profiles find tropopause widths closer to 1 dex in pressure. On the dayside, we are likely not probing deep enough in the atmosphere to constrain the location of the deep temperature structure. Expanding the pressure range over which observations probe will help constrain the precise shape and behavior of the temperature structures in these highly irradiated objects.}

\subsection{Comparison to Ultra-hot Jupiter Phase Curves and Spectra}

\begin{figure}[ht!]
\epsscale{1.2}
\plotone{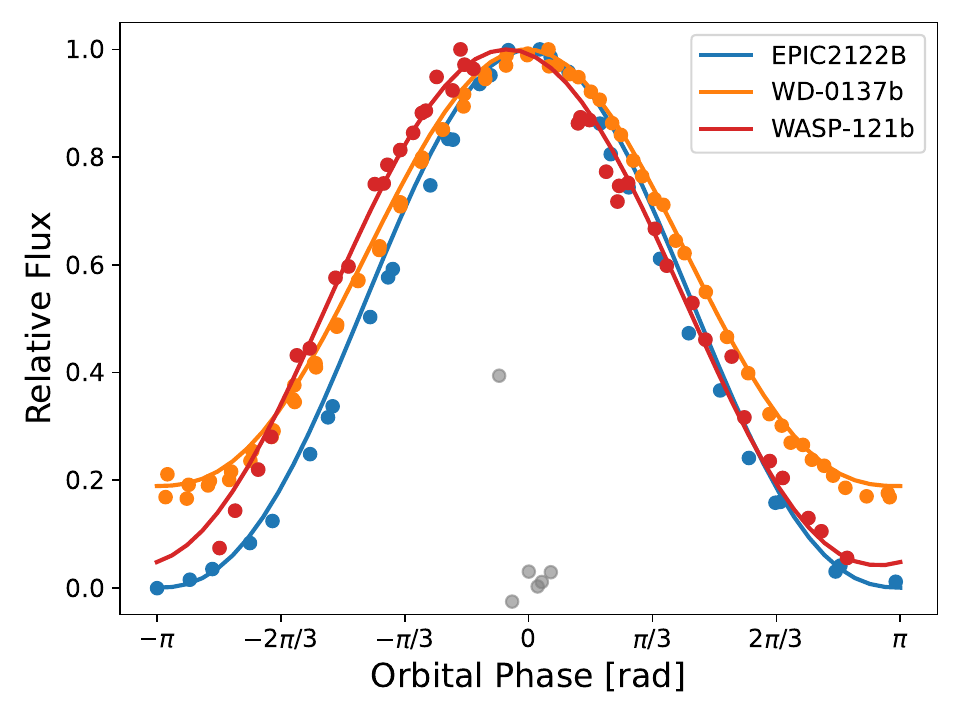}
\caption{Broadband HST/WFC3/G141 phase curve observations (dots) or EPIC-2122B, WD-0137B, and ultra-hot Jupiter WASP-121b. Observations during WASP-121b's transit are colored gray. The lines are fits using the sinusoidal function of Eq.~\ref{eq:etadelta}.
\label{fig:pc_fit_all}}
\end{figure}

In many ways, EPIC-2122B and WD-0137B are analogous to their highly-irradiated, lower-mass main-sequence counterparts: the ultra-hot Jupiters \citep{kitzmann:2018,lothringer:2018b,parmentier:2018}. Ultra-hot Jupiters can reach equilibrium temperatures exceeding 2,000~K with orbital periods on the order of 1 day. However, WD-0137B and EPIC-2122B are different from the population of{ known transiting} ultra-hot Jupiters for a number of important physical reasons, including the{ significantly} inclined viewing geometry, surface gravity that can be about two orders of magnitude higher, the enormous Coriolis forces caused by their even shorter orbital periods (1.13 and 1.93 hours, respectively). The latter two factors lead to relatively small Rossby deformation radii, placing irradiated brown dwarfs in a fundamentally different dynamical regime than ultra-hot Jupiters \citep{tan:2020,lee:2020}. The small Rossby deformation radius of these brown dwarfs is predicted to lead to a narrowing of the equatorial jet and a suppression of the day-to-night heat transfer efficiency \citep{showman:2011}.

In Figure~\ref{fig:pc_fit_all}, we compare the broadband HST/G141 phase curve of EPIC-2122B and WD-0137B to the canonical ultra-hot Jupiter WASP-121b (T$_{eq}=2,450$K, \cite{delrez:2016}) from \cite{mikal-evans:2022}.{ We chose WASP-121b because it has both a phase curve observed with HST/G141 as studied in this work and is representative of similar $\sim 1$ Jupiter-mass ultra-hot planets.} WASP-121b shows a clear eastward offset of $\approx$10$^{\circ}$. The atmosphere of WASP-121b appears to be able to transport heat from the most intense of the incoming irradiation at least several degrees from the sub-stellar point, even in a regime where the magnetic Lorentz force is is expected to be an important source of drag \citep{rogersandtad:2014,beltz:2022}. Offsets are more difficult to measure for EPIC-2122B and WD-0137B due to their non-transiting nature, but \cite{zhou:2022} did not detect an phase offset in WD-0137B using the ephemeris{ measured from the orbital radial velocities} in \cite{longstaff:2017}. Furthermore, \cite{zhou:2022} found that the higher-order sinusoidal components of fits to both objects' phase curve did not show any statistically significant{ relative} offset from the base sinusoidal variation, nor was there any apparent wavelength-dependent offset. 

The overall shape of the phase curve is different for these objects as well. \cite{mikal-evans:2022} found that Equation~\ref{eq:etaphi} was sufficient to describe the partitioning of the dayside and nightside hemispheres for joint phase curve retrievals. However, for both WD-0137B and EPIC-2122B, it was necessary to include the additional free parameter $\delta$ in Eq.~\ref{eq:etadelta}. Because $\delta$ was found to be greater than unity in both brown dwarfs, this implies that the flux from these objects falls off faster than a first-order sinusoid. This is supported by \cite{zhou:2022}'s need for a higher-order sinusoidal fit to the phase curves. {We note that \cite{zhou:2022} found that the GCMs for both objects fit the observed phase curve well compared to a spherical harmonic model and a radiative-convective equilibrium model once the amplitude was scaled to match the observations.}

EPIC-2122B shows very little flux on the nightside, indicating a very large day-night temperature difference, even despite the  contribution to the nightside flux from high-latitudes on the dayside hemisphere due to the inclined viewing geometry of the system. The opposite appears to be true for the even lower inclination WD-0137B, which has a lower day-night relative contrast. We can quantify this by measuring the relative day-night flux contrast $A_F = (F_{\mathrm{day}}-F_{\mathrm{night}})/F_{\mathrm{day}}$, where $F_{\mathrm{day}}$ and $F_{\mathrm{night}}$ are the measured flux on the day and night, respectively. EPIC-2122B has $A_F = 0.98 \pm 0.04$ with WASP-121b having a similar $A_F = 0.95 \pm 0.025$ \citep{mikal-evans:2022}. Meanwhile, WD-0137B's $A_F$ is lower at $0.84 \pm 0.12$

WD-0137B's lower relative day-night flux contrast may in part be because of the amount of dayside still visible at nightside phases, in line with predictions from GCMs of non-transiting planets \citep{malsky:2021}, but we hypothesize that this could also be from the internal temperature, $\sim$1,520~K, which can make up a larger proportion of the overall flux than the more highly-irradiated EPIC-2122B. The much lower mass WASP-121b ($1.157\pm0.07$~M$_J$, \cite{bourrier:2020}) would be expected to have a much lower internal temperature, between about 400 and 600~K \citep{thorngren:2019} and thus might also show a higher relative day-night contrast than WD-0137B. Higher-precision data and wider wavelength coverage where multiple TP profile retrievals will be more likely to be statistically justified could provide a more accurate measurement of the internal temperature, as well as the contribution of the visible dayside component on the nightside. 

Further differences exist between the phase-resolved spectra of WASP-121b compared to the brown dwarfs. EPIC-2122B is significantly hotter than WASP-121b and experiences a higher degree of thermal dissociation of molecules, as well as a greater abundance of H$^-$ compared to WASP-121b. H$^-$ is indeed detected on WASP-121b, but at a lower abundance and only on the dayside (see \cite{mikal-evans:2022}, Figure~3), while EPIC-2122B's spectra are consistent with H$^-$ on the nightside hemisphere (see Figure~\ref{fig:epic2122_abunds}). While WASP-121b has significant H$_2$O emission features on the dayside and absorption on the nightside, EPIC-2122B's spectra remain devoid of H$_2$O features. These differences are well in-line with expectations from EPIC-2122B's higher temperature and reflect the same differences seen between the two brown dwarfs themselves.

Perhaps the more interesting comparison is between WASP-121b and WD-0137B. The dayside spectrum of WASP-121b shows a relatively clear H$_2$O emission feature throughout the entire dayside hemisphere, from phases $\phi = -\frac{\pi}{2}$ to $ \frac{\pi}{2}$. WD-0137B, on the other hand, does not exhibit such a clear feature. While the brightness temperatures shown in Figure~\ref{fig:wd0137_BTs} do exhibit an increase where H$_2$O opacity is at its strongest ($\approx1.45\mu{m}$) that would be consistent with an inversion, all retrievals chose to fit a featureless spectrum for the dayside hemisphere. The cause of this difference could simply be due to the fact that the dayside atmosphere is more isothermal at the photosphere for WD-0137B compared to WASP-121b, perhaps due to the somewhat lower T$_{eq}${ of WD-0137B}. However, the dayside retrievals did constrain a relatively strong temperature inversion with the information from the H$^-$ opacity (see Figure~\ref{fig:wd0137_TPs}). Another explanation might again involve the inclined viewing geometry: any H$_2$O emission on the dayside was diluted by H$_2$O absorption from the still-visible nightside and/or from the less directly irradiated high-latitudes. The precision of the data did not justify two TP profiles to represent these components in the retrieval, but future higher-precision data may be capable of distinguishing these components. 

We note that WD-0137B does have atomic lines in emission \citep{longstaff:2017}, confirming a temperature inversion on the dayside and temperatures warm enough for atomic H, He, Na, Mg, Si, K, Ca, Ti, and Fe. While many other ultra-hot Jupiters also show atomic lines in emission \citep[e.g.,][]{pino:2020}, and while WASP-121b shows a plethora of atomic lines in transmission \citep{sing:2019,cabot:2020,borsa:2020,gibson:2020,merritt:2021,azevedosilva:2022,maguire:2023,seidel:2023}, there have been no atomic emission lines published yet for WASP-121b to our knowledge.

The differences between WASP-121b and WD-0137B extend to the nightside as well. WASP-121b's spectra remain fairly featureless until near the anti-stellar point where a clear H$_2$O absorption feature is evident. In contrast, WD-0137B exhibits obvious H$_2$O absorption at pre-dawn and evening phases.
As soon as the atmosphere is no longer being externally forced by the irradiation to the temperatures that induce the temperature inversion, WD-0137B's atmosphere relaxes back to a non-inverted structure. More formally, this shows the radiative timescale at which the object can cool itself is much less than the advective timescale at which heat can be transported. 

This is rather unlike EPIC-2122B, which does remain hot enough for the temperature inversion to remain at phases on the nightside hemisphere. The difference between WD-0137B and EPIC-2122B is likely the more extreme temperature to which EPIC-2122B is irradiated. When combined with the 56$^{\circ}$ inclination of EPIC-2122B's viewing geometry: high-latitudes on the dayside hemisphere are always visible for EPIC-2122B and thus the extremely irradiated and inverted portions of the atmosphere might still be influencing the shape of the spectrum, even at evening and pre-dawn phases. While WD-0137B's orbital inclination is even lower (and thus the observered nadir latitude is higher), its high-latitude dayside atmosphere does not appear to be hot enough to dominate the nightside spectra. 

Looking toward the future, the phase curve of WASP-121b has been observed with JWST/NIRSpec G395H \citep{mikal-evans:2023} and with JWST/NIRISS/SOSS (JWST-GTO-1201) for complete near-infrared coverage from 0.6-5$\mu{m}$. However, corresponding phase curves of WD-0137B, EPIC-2122B, or any irradiated brown dwarf have yet to be observed with JWST. {Fortunately, JWST-GO-4967 is a recently approved Cycle 3 program to observe five WD-BD systems, including WD-0137, with NIRSpec/PRISM (0.6-5.2 $\mu$m).} As we have found, these irradiated brown dwarfs exhibit different characteristics compared to lower-mass objects when it comes to their emission spectra and atmospheric circulation. Expanding this comparison to ultra-hot Jupiters to longer wavelengths with JWST will reveal how carbon-bearing species like CO behave and how differences in the circulation pattern hold for a wider range of atmospheric pressures. Furthermore, mysteries still remain around the anomalously bright $K_s$ and \textit{Spitzer} 4.5$\mu{m}$ fluxes for WD-0137B \citep{casewell:2015}{, for which H$_3^+$ emission has been offered as an explanation, but could also be emission from carbon species, including CO}.

Similarly, expanding the sample of observed irradiated brown dwarfs is also important. As seen here, orbital inclination can consequential when it comes to interpreting the observed spectrum. By observing more objects with different inclinations, we can better understand such effects while learning about the nature of these objects' high-latitude atmosphere. An example would be observations of SDSS J1205B (a.k.a. EPIC-20128311, \cite{parsons:2017}), which is exceedingly similar to EPIC-2122B except that it is eclipsing and thus has an orbital inclination near $90^{\circ}$. A similar strategy would help us understand the influence of orbital period/rotation rate, UV irradiation, etc.{, on the global circulation.}


\section{Conclusion}

We have performed a comprehensive suite of atmospheric retrievals on the HST/WFC3 G141 spectroscopic light curves of two highly irradiated brown dwarfs, WD-0137B and EPIC-2122B, that both have white dwarfs as host stars. While these systems are analogous to hot and ultra-hot Jupiters in terms of total irradiation and atmospheric chemistry, these systems are also unique when it comes to their rotation rate, surface gravity, extreme UV irradiation, and viewing geometry. 

The high signal-to-noise of the observations enables precise phase-resolved constraints on the atmospheric abundances of H$_2$O and H$^-$ (through the e$^-$ density), as well as the temperature structure. The phase curve of WD-0137B reveal an inverted atmosphere dominated by H$^-$ opacity on the dayside that, once on the nightside hemisphere, quickly transitions to a non-inverted atmosphere with strong H$_2$O absorption. The much hotter EPIC-2122B, on the other, exhibits an inverted atmosphere onto the nightside, with the continuous opacity of H$^-$ dominating the spectrum at every longitude except the relatively unconstrained anti-stellar point. 

We ran a variety of retrievals, including fiducial longitudinally independent retrievals, retrievals with non-vertically-uniform atmospheric abundances, and retrievals with dual-temperature structures. For the dayside of WD-0137B, the fiducial retrievals were favored by the Bayesian Information Criterion (BIC), with positive evidence for a dayside dilution factor to account for a flux-dominating hot spot. For the dayside of EPIC-2122B, retrievals including a dilution factor and dual-TP profiles were justified, with the former being favored due to its simplicity, in line with conclusions from exoplanet retrievals \citep{taylor:2020}. We also ran joint retrievals to all phases simultaneously, and found a linearly-combined dayside and nightside atmosphere provided an adequate fit to the data for both objects. 

We then compared these results to GCM predictions for these two objects and to similar observations of the ultra-hot Jupiter WASP-121b. These results show the rich insight these phase-resolved spectroscopic observations can give into the atmospheric composition, temperatures structure, and circulation of these unique atmospheres. {We look forward to the new observations that JWST will provide in JWST-GO-4967, which will extend these phase curves to longer wavelengths and reveal the behavior of species beyond H$_2$O and H$^-$ to provide a more comprehensive look at the longitudinally-resolved emission of these extreme objects.}

\begin{acknowledgments}
We thank the anonymous referee for comments and suggestions that improved the manuscript. We thank David Sing, Travis Barman, and Peter Hauschildt for the use of computing facilities used to carry out this analysis. The authors gratefully acknowledge the computing time granted by the Resource Allocation Board and provided on the supercomputer Lise and Emmy at NHR@ZIB and NHR@Göttingen as part of the NHR infrastructure. The calculations for this research were conducted with computing resources under the project HHP00051. All {\it HST} data used in this paper can be found in MAST: \dataset[10.17909/bsny-4f21]{http://dx.doi.org/10.17909/bsny-4f21}. This work is based on observations with the NASA/ESA Hubble Space Telescope obtained at the Space Telescope Science Institute, which is operated by the Association of Universities for Research in Astronomy, Incorporated, under NASA contract NAS5-26555. Support for Program number HST-GO-15947 and HST-AR-16142 and was provided through a grant from the STScI under NASA contract NAS5-26555.
\end{acknowledgments}

%

\vspace{5mm}
\facilities{HST (WFC3)}


\software{astropy \citep{astropy:2018}, matplotlib \citep{hunter:2007}, NumPy \citep{numpy:2020}, iPython \citep{perez:2007}, PHOENIX \citep{hauschildt:1997}
          }


\clearpage
\bibliographystyle{aasjournal}

\begin{thebibliography}{}
\expandafter\ifx\csname natexlab\endcsname\relax\def\natexlab#1{#1}\fi
\providecommand{\url}[1]{\href{#1}{#1}}
\providecommand{\dodoi}[1]{doi:~\href{http://doi.org/#1}{\nolinkurl{#1}}}
\providecommand{\doeprint}[1]{\href{http://ascl.net/#1}{\nolinkurl{http://ascl.net/#1}}}
\providecommand{\doarXiv}[1]{\href{https://arxiv.org/abs/#1}{\nolinkurl{https://arxiv.org/abs/#1}}}

\bibitem[{{Akaike}(1974)}]{akaike:1974}
{Akaike}, H. 1974, IEEE Transactions on Automatic Control, 19, 716

\bibitem[{{Amaro} {et~al.}(2023){Amaro}, {Apai}, {Zhou}, {Lew}, {Casewell}, {Mayorga}, {Marley}, {Tan}, {Lothringer}, {Parmentier}, \& {Barman}}]{amaro:2023}
{Amaro}, R.~C., {Apai}, D., {Zhou}, Y., {et~al.} 2023, \apj, 948, 129, \dodoi{10.3847/1538-4357/acbfb3}

\bibitem[{{Apai} {et~al.}(2019){Apai}, {Marley}, {Showman}, {Xu}, \& {Zhou}}]{apai:15947}
{Apai}, D., {Marley}, M.~S., {Showman}, A., {Xu}, S., \& {Zhou}, Y. 2019, {Dancing with the Dwarfs: Very High Quality Spatial and Spectral Maps of Hot Jupiters Proxies}, HST Proposal. Cycle 27, ID. \#15947

\bibitem[{{Arcangeli} {et~al.}(2018){Arcangeli}, {D{\'e}sert}, {Line}, {Bean}, {Parmentier}, {Stevenson}, {Kreidberg}, {Fortney}, {Mansfield}, \& {Showman}}]{arcangeli:2018}
{Arcangeli}, J., {D{\'e}sert}, J.-M., {Line}, M.~R., {et~al.} 2018, \apjl, 855, L30, \dodoi{10.3847/2041-8213/aab272}

\bibitem[{{Arcangeli} {et~al.}(2019){Arcangeli}, {Desert}, {Parmentier}, {Stevenson}, {Bean}, {Line}, {Kreidberg}, {Fortney}, \& {Showman}}]{arcangeli:2019}
{Arcangeli}, J., {Desert}, J.-M., {Parmentier}, V., {et~al.} 2019, arXiv e-prints.
\newblock \doarXiv{1904.02069}

\bibitem[{{Astropy Collaboration} {et~al.}(2018){Astropy Collaboration}, {Price-Whelan}, {Sip{\H{o}}cz}, {G{\"u}nther}, {Lim}, {Crawford}, {Conseil}, {Shupe}, {Craig}, {Dencheva}, {Ginsburg}, {VanderPlas}, {Bradley}, {P{\'e}rez-Su{\'a}rez}, {de Val-Borro}, {Aldcroft}, {Cruz}, {Robitaille}, {Tollerud}, {Ardelean}, {Babej}, {Bach}, {Bachetti}, {Bakanov}, {Bamford}, {Barentsen}, {Barmby}, {Baumbach}, {Berry}, {Biscani}, {Boquien}, {Bostroem}, {Bouma}, {Brammer}, {Bray}, {Breytenbach}, {Buddelmeijer}, {Burke}, {Calderone}, {Cano Rodr{\'\i}guez}, {Cara}, {Cardoso}, {Cheedella}, {Copin}, {Corrales}, {Crichton}, {D'Avella}, {Deil}, {Depagne}, {Dietrich}, {Donath}, {Droettboom}, {Earl}, {Erben}, {Fabbro}, {Ferreira}, {Finethy}, {Fox}, {Garrison}, {Gibbons}, {Goldstein}, {Gommers}, {Greco}, {Greenfield}, {Groener}, {Grollier}, {Hagen}, {Hirst}, {Homeier}, {Horton}, {Hosseinzadeh}, {Hu}, {Hunkeler}, {Ivezi{\'c}}, {Jain}, {Jenness}, {Kanarek}, {Kendrew}, {Kern}, {Kerzendorf}, {Khvalko}, {King}, {Kirkby}, {Kulkarni},
  {Kumar}, {Lee}, {Lenz}, {Littlefair}, {Ma}, {Macleod}, {Mastropietro}, {McCully}, {Montagnac}, {Morris}, {Mueller}, {Mumford}, {Muna}, {Murphy}, {Nelson}, {Nguyen}, {Ninan}, {N{\"o}the}, {Ogaz}, {Oh}, {Parejko}, {Parley}, {Pascual}, {Patil}, {Patil}, {Plunkett}, {Prochaska}, {Rastogi}, {Reddy Janga}, {Sabater}, {Sakurikar}, {Seifert}, {Sherbert}, {Sherwood-Taylor}, {Shih}, {Sick}, {Silbiger}, {Singanamalla}, {Singer}, {Sladen}, {Sooley}, {Sornarajah}, {Streicher}, {Teuben}, {Thomas}, {Tremblay}, {Turner}, {Terr{\'o}n}, {van Kerkwijk}, {de la Vega}, {Watkins}, {Weaver}, {Whitmore}, {Woillez}, {Zabalza}, \& {Astropy Contributors}}]{astropy:2018}
{Astropy Collaboration}, {Price-Whelan}, A.~M., {Sip{\H{o}}cz}, B.~M., {et~al.} 2018, \aj, 156, 123, \dodoi{10.3847/1538-3881/aabc4f}

\bibitem[{{Azevedo Silva} {et~al.}(2022){Azevedo Silva}, {Demangeon}, {Santos}, {Allart}, {Borsa}, {Cristo}, {Esparza-Borges}, {Seidel}, {Palle}, {Sousa}, {Tabernero}, {Zapatero Osorio}, {Cristiani}, {Pepe}, {Rebolo}, {Adibekyan}, {Alibert}, {Barros}, {Bouchy}, {Bourrier}, {Lo Curto}, {Di Marcantonio}, {D'Odorico}, {Ehrenreich}, {Figueira}, {Gonz{\'a}lez Hern{\'a}ndez}, {Lovis}, {Martins}, {Mehner}, {Micela}, {Molaro}, {Mounzer}, {Nunes}, {Sozzetti}, {Su{\'a}rez Mascare{\~n}o}, \& {Udry}}]{azevedosilva:2022}
{Azevedo Silva}, T., {Demangeon}, O.~D.~S., {Santos}, N.~C., {et~al.} 2022, \aap, 666, L10, \dodoi{10.1051/0004-6361/202244489}

\bibitem[{{Barber} {et~al.}(2006){Barber}, {Tennyson}, {Harris}, \& {Tolchenov}}]{barber:2006}
{Barber}, R.~J., {Tennyson}, J., {Harris}, G.~J., \& {Tolchenov}, R.~N. 2006, \mnras, 368, 1087, \dodoi{10.1111/j.1365-2966.2006.10184.x}

\bibitem[{{Barman} {et~al.}(2001){Barman}, {Hauschildt}, \& {Allard}}]{barman:2001}
{Barman}, T.~S., {Hauschildt}, P.~H., \& {Allard}, F. 2001, \apj, 556, 885, \dodoi{10.1086/321610}

\bibitem[{{Beatty} {et~al.}(2017){Beatty}, {Madhusudhan}, {Pogge}, {Chung}, {Bierlya}, {Gaudi}, \& {Latham}}]{beatty:2017b}
{Beatty}, T.~G., {Madhusudhan}, N., {Pogge}, R., {et~al.} 2017, \aj, 154, 242, \dodoi{10.3847/1538-3881/aa94cf}

\bibitem[{Beatty {et~al.}(2018)Beatty, Morley, Curtis, Burrows, Davenport, \& Montet}]{beatty:2018b}
Beatty, T.~G., Morley, C.~V., Curtis, J.~L., {et~al.} 2018.
\newblock \doarXiv{1807.11500}

\bibitem[{{Bell} \& {Cowan}(2018)}]{bell:2018}
{Bell}, T.~J., \& {Cowan}, N.~B. 2018, \apjl, 857, L20, \dodoi{10.3847/2041-8213/aabcc8}

\bibitem[{{Beltz} {et~al.}(2021){Beltz}, {Rauscher}, {Brogi}, \& {Kempton}}]{beltz:2021}
{Beltz}, H., {Rauscher}, E., {Brogi}, M., \& {Kempton}, E. M.~R. 2021, \aj, 161, 1, \dodoi{10.3847/1538-3881/abb67b}

\bibitem[{{Beltz} {et~al.}(2022){Beltz}, {Rauscher}, {Roman}, \& {Guilliat}}]{beltz:2022}
{Beltz}, H., {Rauscher}, E., {Roman}, M.~T., \& {Guilliat}, A. 2022, \aj, 163, 35, \dodoi{10.3847/1538-3881/ac3746}

\bibitem[{{Borsa} {et~al.}(2020){Borsa}, {Allart}, {Casasayas-Barris}, {Tabernero}, {Zapatero Osorio}, {Cristiani}, {Pepe}, {Rebolo}, {Santos}, {Adibekyan}, {Bourrier}, {Demangeon}, {Ehrenreich}, {Pall{\'e}}, {Sousa}, {Lillo-Box}, {Lovis}, {Micela}, {Oshagh}, {Poretti}, {Sozzetti}, {Allende Prieto}, {Alibert}, {Amate}, {Benz}, {Bouchy}, {Cabral}, {Dekker}, {D'Odorico}, {Di Marcantonio}, {Figueira}, {Genova Santos}, {Gonz{\'a}lez Hern{\'a}ndez}, {Lo Curto}, {Manescau}, {Martins}, {M{\'e}gevand}, {Mehner}, {Molaro}, {Nunes}, {Riva}, {Su{\'a}rez Mascare{\~n}o}, {Udry}, \& {Zerbi}}]{borsa:2020}
{Borsa}, F., {Allart}, R., {Casasayas-Barris}, N., {et~al.} 2020, arXiv e-prints, arXiv:2011.01245.
\newblock \doarXiv{2011.01245}

\bibitem[{{Bourrier} {et~al.}(2020){Bourrier}, {Ehrenreich}, {Lendl}, {Cretignier}, {Allart}, {Dumusque}, {Cegla}, {Suarez-Mascareno}, {Wyttenbach}, {Hoeijmakers}, {Melo}, {Kuntzer}, {Astudillo-Defru}, {Giles}, {Heng}, {Kitzmann}, {Lavie}, {Lovis}, {Murgas}, {Nascimbeni}, {Pepe}, {Pino}, {Segransan}, \& {Udry}}]{bourrier:2020}
{Bourrier}, V., {Ehrenreich}, D., {Lendl}, M., {et~al.} 2020, arXiv e-prints, arXiv:2001.06836.
\newblock \doarXiv{2001.06836}

\bibitem[{{Burleigh} {et~al.}(2006){Burleigh}, {Hogan}, {Dobbie}, {Napiwotzki}, \& {Maxted}}]{burleigh:2006}
{Burleigh}, M.~R., {Hogan}, E., {Dobbie}, P.~D., {Napiwotzki}, R., \& {Maxted}, P.~F.~L. 2006, \mnras, 373, L55, \dodoi{10.1111/j.1745-3933.2006.00242.x}

\bibitem[{{Cabot} {et~al.}(2020){Cabot}, {Madhusudhan}, {Welbanks}, {Piette}, \& {Gandhi}}]{cabot:2020}
{Cabot}, S. H.~C., {Madhusudhan}, N., {Welbanks}, L., {Piette}, A., \& {Gandhi}, S. 2020, \mnras, 494, 363, \dodoi{10.1093/mnras/staa748}

\bibitem[{{Casewell} {et~al.}(2015){Casewell}, {Lawrie}, {Maxted}, {Marley}, {Fortney}, {Rimmer}, {Littlefair}, {Wynn}, {Burleigh}, \& {Helling}}]{casewell:2015}
{Casewell}, S.~L., {Lawrie}, K.~A., {Maxted}, P.~F.~L., {et~al.} 2015, \mnras, 447, 3218, \dodoi{10.1093/mnras/stu2721}

\bibitem[{{Casewell} {et~al.}(2018){Casewell}, {Braker}, {Parsons}, {Hermes}, {Burleigh}, {Belardi}, {Chaushev}, {Finch}, {Roy}, {Littlefair}, {Goad}, \& {Dennihy}}]{casewell:2018}
{Casewell}, S.~L., {Braker}, I.~P., {Parsons}, S.~G., {et~al.} 2018, \mnras, 476, 1405, \dodoi{10.1093/mnras/sty245}

\bibitem[{{Coulombe} {et~al.}(2023){Coulombe}, {Benneke}, {Challener}, {Piette}, {Wiser}, {Mansfield}, {MacDonald}, {Beltz}, {Feinstein}, {Radica}, {Savel}, {Dos Santos}, {Bean}, {Parmentier}, {Wong}, {Rauscher}, {Komacek}, {Kempton}, {Tan}, {Hammond}, {Lewis}, {Line}, {Lee}, {Shivkumar}, {Crossfield}, {Nixon}, {Rackham}, {Wakeford}, {Welbanks}, {Zhang}, {Batalha}, {Berta-Thompson}, {Changeat}, {D{\'e}sert}, {Espinoza}, {Goyal}, {Harrington}, {Knutson}, {Kreidberg}, {L{\'o}pez-Morales}, {Shporer}, {Sing}, {Stevenson}, {Aggarwal}, {Ahrer}, {Alam}, {Bell}, {Blecic}, {Caceres}, {Carter}, {Casewell}, {Crouzet}, {Cubillos}, {Decin}, {Fortney}, {Gibson}, {Heng}, {Henning}, {Iro}, {Kendrew}, {Lagage}, {Leconte}, {Lendl}, {Lothringer}, {Mancini}, {Mikal-Evans}, {Molaverdikhani}, {Nikolov}, {Ohno}, {Palle}, {Piaulet}, {Redfield}, {Roy}, {Tsai}, {Venot}, \& {Wheatley}}]{coulombe:2023}
{Coulombe}, L.-P., {Benneke}, B., {Challener}, R., {et~al.} 2023, arXiv e-prints, arXiv:2301.08192.
\newblock \doarXiv{2301.08192}

\bibitem[{{Curtis} {et~al.}(2016){Curtis}, {Vanderburg}, {Montet}, {Beatty}, {Bieryla}, {Cargile}, {Kraus}, {Latham}, {Mann}, {Nofi}, {Rizzuto}, {Saar}, \& {Wright}}]{curtis:2016}
{Curtis}, J., {Vanderburg}, A., {Montet}, B., {et~al.} 2016, in 19th Cambridge Workshop on Cool Stars, Stellar Systems, and the Sun (CS19), Cambridge Workshop on Cool Stars, Stellar Systems, and the Sun, 95, \dodoi{10.5281/zenodo.58758}

\bibitem[{{Delrez} {et~al.}(2016){Delrez}, {Santerne}, {Almenara}, {Anderson}, {Collier-Cameron}, {D{\'\i}az}, {Gillon}, {Hellier}, {Jehin}, {Lendl}, {Maxted}, {Neveu-VanMalle}, {Pepe}, {Pollacco}, {Queloz}, {S{\'e}gransan}, {Smalley}, {Smith}, {Triaud}, {Udry}, {Van Grootel}, \& {West}}]{delrez:2016}
{Delrez}, L., {Santerne}, A., {Almenara}, J.~M., {et~al.} 2016, \mnras, 458, 4025.
\newblock \doarXiv{1506.02471}

\bibitem[{{Dupuy} \& {Liu}(2012)}]{dupuy:2012}
{Dupuy}, T.~J., \& {Liu}, M.~C. 2012, \apjs, 201, 19, \dodoi{10.1088/0067-0049/201/2/19}

\bibitem[{{Farihi} \& {Christopher}(2004)}]{farihi:2004}
{Farihi}, J., \& {Christopher}, M. 2004, \aj, 128, 1868, \dodoi{10.1086/423919}

\bibitem[{{Feng} {et~al.}(2020){Feng}, {Line}, \& {Fortney}}]{feng:2020}
{Feng}, Y.~K., {Line}, M.~R., \& {Fortney}, J.~J. 2020, \aj, 160, 137, \dodoi{10.3847/1538-3881/aba8f9}

\bibitem[{{Gelman} \& {Rubin}(1992)}]{gelman:1992}
{Gelman}, A., \& {Rubin}, D.~B. 1992, Statistical Science, 7, 457, \dodoi{10.1214/ss/1177011136}

\bibitem[{{Gibson} {et~al.}(2020){Gibson}, {Merritt}, {Nugroho}, {Cubillos}, {de Mooij}, {Mikal-Evans}, {Fossati}, {Lothringer}, {Nikolov}, {Sing}, {Spake}, {Watson}, \& {Wilson}}]{gibson:2020}
{Gibson}, N.~P., {Merritt}, S., {Nugroho}, S.~K., {et~al.} 2020, \mnras, 493, 2215, \dodoi{10.1093/mnras/staa228}

\bibitem[{Harris {et~al.}(2020)Harris, Millman, van~der Walt, Gommers, Virtanen, Cournapeau, Wieser, Taylor, Berg, Smith, Kern, Picus, Hoyer, van Kerkwijk, Brett, Haldane, FernÃ¡ndez~del RÃ­o, Wiebe, Peterson, GÃ©rard-Marchant, Sheppard, Reddy, Weckesser, Abbasi, Gohlke, \& Oliphant}]{numpy:2020}
Harris, C.~R., Millman, K.~J., van~der Walt, S.~J., {et~al.} 2020, Nature, 585, 357â€“362, \dodoi{10.1038/s41586-020-2649-2}

\bibitem[{{Hauschildt} {et~al.}(1999){Hauschildt}, {Allard}, \& {Baron}}]{hauschildt:1999}
{Hauschildt}, P.~H., {Allard}, F., \& {Baron}, E. 1999, \apj, 512, 377, \dodoi{10.1086/306745}

\bibitem[{{Hauschildt} {et~al.}(1997){Hauschildt}, {Baron}, \& {Allard}}]{hauschildt:1997}
{Hauschildt}, P.~H., {Baron}, E., \& {Allard}, F. 1997, \apj, 483, 390, \dodoi{10.1086/304233}

\bibitem[{Hunter(2007)}]{hunter:2007}
Hunter, J.~D. 2007, Computing in Science \& Engineering, 9, 90, \dodoi{10.1109/MCSE.2007.55}

\bibitem[{{John}(1988)}]{john:1988}
{John}, T.~L. 1988, \aap, 193, 189

\bibitem[{{Kitzmann} {et~al.}(2018){Kitzmann}, {Heng}, {Rimmer}, {Hoeijmakers}, {Tsai}, {Malik}, {Lendl}, {Deitrick}, \& {Demory}}]{kitzmann:2018}
{Kitzmann}, D., {Heng}, K., {Rimmer}, P.~B., {et~al.} 2018, \apj, 863, 183, \dodoi{10.3847/1538-4357/aace5a}

\bibitem[{{Koester}(2010)}]{koester:2010}
{Koester}, D. 2010, Memorie della Societa Astronomica Italiana, 81, 921

\bibitem[{{Komacek} \& {Tan}(2018)}]{tad:2018}
{Komacek}, T.~D., \& {Tan}, X. 2018, Research Notes of the American Astronomical Society, 2, 36, \dodoi{10.3847/2515-5172/aac5e7}

\bibitem[{{Lee} {et~al.}(2020){Lee}, {Casewell}, {Chubb}, {Hammond}, {Tan}, {Tsai}, \& {Pierrehumbert}}]{lee:2020}
{Lee}, E.~K.~H., {Casewell}, S.~L., {Chubb}, K.~L., {et~al.} 2020, \mnras, 496, 4674, \dodoi{10.1093/mnras/staa1882}

\bibitem[{{Lew} {et~al.}(2022){Lew}, {Apai}, {Zhou}, {Marley}, {Mayorga}, {Tan}, {Parmentier}, {Casewell}, \& {Xu (许偲艺)}}]{lew:2022}
{Lew}, B. W.~P., {Apai}, D., {Zhou}, Y., {et~al.} 2022, \aj, 163, 8, \dodoi{10.3847/1538-3881/ac3001}

\bibitem[{{Longstaff} {et~al.}(2017){Longstaff}, {Casewell}, {Wynn}, {Maxted}, \& {Helling}}]{longstaff:2017}
{Longstaff}, E.~S., {Casewell}, S.~L., {Wynn}, G.~A., {Maxted}, P.~F.~L., \& {Helling}, C. 2017, \mnras, 471, 1728, \dodoi{10.1093/mnras/stx1786}

\bibitem[{{Lothringer} {et~al.}(2018){Lothringer}, {Barman}, \& {Koskinen}}]{lothringer:2018b}
{Lothringer}, J.~D., {Barman}, T., \& {Koskinen}, T. 2018, \apj, 866, 27, \dodoi{10.3847/1538-4357/aadd9e}

\bibitem[{{Lothringer} \& {Barman}(2020)}]{lothringer:2020a}
{Lothringer}, J.~D., \& {Barman}, T.~S. 2020, \aj, 159, 289, \dodoi{10.3847/1538-3881/ab8d33}

\bibitem[{{Lothringer} \& {Casewell}(2020)}]{lothringer:2020c}
{Lothringer}, J.~D., \& {Casewell}, S.~L. 2020, arXiv e-prints, arXiv:2010.14319.
\newblock \doarXiv{2010.14319}

\bibitem[{{Madhusudhan} \& {Seager}(2009)}]{madhusudhan:2009}
{Madhusudhan}, N., \& {Seager}, S. 2009, \apj, 707, 24, \dodoi{10.1088/0004-637X/707/1/24}

\bibitem[{{Maguire} {et~al.}(2023){Maguire}, {Gibson}, {Nugroho}, {Ramkumar}, {Fortune}, {Merritt}, \& {de Mooij}}]{maguire:2023}
{Maguire}, C., {Gibson}, N.~P., {Nugroho}, S.~K., {et~al.} 2023, \mnras, 519, 1030, \dodoi{10.1093/mnras/stac3388}

\bibitem[{{Majeau} {et~al.}(2012){Majeau}, {Agol}, \& {Cowan}}]{majeau:2012}
{Majeau}, C., {Agol}, E., \& {Cowan}, N.~B. 2012, \apjl, 747, L20, \dodoi{10.1088/2041-8205/747/2/L20}

\bibitem[{{Malsky} {et~al.}(2021){Malsky}, {Rauscher}, {Kempton}, {Roman}, {Long}, \& {Harada}}]{malsky:2021}
{Malsky}, I., {Rauscher}, E., {Kempton}, E. M.~R., {et~al.} 2021, \apj, 923, 62, \dodoi{10.3847/1538-4357/ac2a2a}

\bibitem[{{Mansfield} {et~al.}(2019){Mansfield}, {Bean}, {Stevenson}, {Komacek}, {Bell}, {Tan}, {Malik}, {Beatty}, {Wong}, {Cowan}, {Dang}, {D{\'e}sert}, {Fortney}, {Gaudi}, {Keating}, {Kempton}, {Kreidberg}, {Line}, {Parmentier}, {Stassun}, {Swain}, \& {Zellem}}]{mansfield:2019b}
{Mansfield}, M., {Bean}, J.~L., {Stevenson}, K.~B., {et~al.} 2019, arXiv e-prints, arXiv:1910.01567.
\newblock \doarXiv{1910.01567}

\bibitem[{{Maxted} {et~al.}(2006){Maxted}, {Napiwotzki}, {Dobbie}, \& {Burleigh}}]{maxted:2006}
{Maxted}, P.~F.~L., {Napiwotzki}, R., {Dobbie}, P.~D., \& {Burleigh}, M.~R. 2006, \nat, 442, 543, \dodoi{10.1038/nature04987}

\bibitem[{{Merritt} {et~al.}(2021){Merritt}, {Gibson}, {Nugroho}, {de Mooij}, {Hooton}, {Lothringer}, {Matthews}, {Mikal-Evans}, {Nikolov}, {Sing}, \& {Watson}}]{merritt:2021}
{Merritt}, S.~R., {Gibson}, N.~P., {Nugroho}, S.~K., {et~al.} 2021, \mnras, \dodoi{10.1093/mnras/stab1878}

\bibitem[{{Mikal-Evans} {et~al.}(2022){Mikal-Evans}, {Sing}, {Barstow}, {Kataria}, {Goyal}, {Lewis}, {Taylor}, {Mayne}, {Daylan}, {Wakeford}, {Marley}, \& {Spake}}]{mikal-evans:2022}
{Mikal-Evans}, T., {Sing}, D.~K., {Barstow}, J.~K., {et~al.} 2022, Nature Astronomy, \dodoi{10.1038/s41550-021-01592-w}

\bibitem[{{Mikal-Evans} {et~al.}(2023){Mikal-Evans}, {Sing}, {Dong}, {Foreman-Mackey}, {Kataria}, {Barstow}, {Goyal}, {Lewis}, {Lothringer}, {Mayne}, {Wakeford}, {Christie}, \& {Rustamkulov}}]{mikal-evans:2023}
{Mikal-Evans}, T., {Sing}, D.~K., {Dong}, J., {et~al.} 2023, \apjl, 943, L17, \dodoi{10.3847/2041-8213/acb049}

\bibitem[{{Nowak} {et~al.}(2017){Nowak}, {Palle}, {Gandolfi}, {Dai}, {Lanza}, {Hirano}, {Barrag{\'a}n}, {Fukui}, {Bruntt}, {Endl}, {Cochran}, {Prada Moroni}, {Prieto-Arranz}, {Kiilerich}, {Nespral}, {Hatzes}, {Albrecht}, {Deeg}, {Winn}, {Yu}, {Kuzuhara}, {Grziwa}, {Smith}, {Guenther}, {Van Eylen}, {Csizmadia}, {Fridlund}, {Cabrera}, {Eigm{\"u}ller}, {Erikson}, {Korth}, {Narita}, {P{\"a}tzold}, {Rauer}, \& {Ribas}}]{nowak:2017}
{Nowak}, G., {Palle}, E., {Gandolfi}, D., {et~al.} 2017, \aj, 153, 131, \dodoi{10.3847/1538-3881/aa5cb6}

\bibitem[{{Parmentier} \& {Guillot}(2014)}]{parmentier:2014}
{Parmentier}, V., \& {Guillot}, T. 2014, \aap, 562, A133, \dodoi{10.1051/0004-6361/201322342}

\bibitem[{{Parmentier} {et~al.}(2018){Parmentier}, {Line}, {Bean}, {Mansfield}, {Kreidberg}, {Lupu}, {Visscher}, {D{\'e}sert}, {Fortney}, {Deleuil}, {Arcangeli}, {Showman}, \& {Marley}}]{parmentier:2018}
{Parmentier}, V., {Line}, M.~R., {Bean}, J.~L., {et~al.} 2018, \aap, 617, A110, \dodoi{10.1051/0004-6361/201833059}

\bibitem[{{Parsons} {et~al.}(2017){Parsons}, {Hermes}, {Marsh}, {G{\"a}nsicke}, {Tremblay}, {Littlefair}, {Sahman}, {Ashley}, {Green}, {Rattanasoon}, {Dhillon}, {Burleigh}, {Casewell}, {Buckley}, {Braker}, {Irawati}, {Dennihy}, {Rodr{\'\i}guez-Gil}, {Winget}, {Winget}, {Bell}, \& {Kilic}}]{parsons:2017}
{Parsons}, S.~G., {Hermes}, J.~J., {Marsh}, T.~R., {et~al.} 2017, \mnras, 471, 976, \dodoi{10.1093/mnras/stx1610}

\bibitem[{Perez \& Granger(2007)}]{perez:2007}
Perez, F., \& Granger, B.~E. 2007, Computing in Science \& Engineering, 9, 21, \dodoi{10.1109/MCSE.2007.53}

\bibitem[{{Pino} {et~al.}(2020){Pino}, {D{\'e}sert}, {Brogi}, {Malavolta}, {Wyttenbach}, {Line}, {Hoeijmakers}, {Fossati}, {Bonomo}, {Nascimbeni}, {Panwar}, {Affer}, {Benatti}, {Biazzo}, {Bignamini}, {Borsa}, {Carleo}, {Claudi}, {Cosentino}, {Covino}, {Damasso}, {Desidera}, {Giacobbe}, {Harutyunyan}, {Lanza}, {Leto}, {Maggio}, {Maldonado}, {Mancini}, {Micela}, {Molinari}, {Pagano}, {Piotto}, {Poretti}, {Rainer}, {Scandariato}, {Sozzetti}, {Allart}, {Borsato}, {Bruno}, {Di Fabrizio}, {Ehrenreich}, {Fiorenzano}, {Frustagli}, {Lavie}, {Lovis}, {Magazz{\`u}}, {Nardiello}, {Pedani}, \& {Smareglia}}]{pino:2020}
{Pino}, L., {D{\'e}sert}, J.-M., {Brogi}, M., {et~al.} 2020, \apjl, 894, L27, \dodoi{10.3847/2041-8213/ab8c44}

\bibitem[{{Rogers} \& {Komacek}(2014)}]{rogersandtad:2014}
{Rogers}, T.~M., \& {Komacek}, T.~D. 2014, \apj, 794, 132, \dodoi{10.1088/0004-637X/794/2/132}

\bibitem[{Schwarz(1978)}]{schwarz:1978}
Schwarz, G. 1978, Annals of Statistics, 6, 461

\bibitem[{{Seidel} {et~al.}(2023){Seidel}, {Borsa}, {Pino}, {Ehrenreich}, {Stangret}, {Zapatero Osorio}, {Palle}, {Alibert}, {Allart}, {Bourrier}, {Di Marcantonio}, {Figueira}, {Gonz{\'a}lez Hern{\'a}ndez}, {Lillo-Box}, {Lovis}, {Martins}, {Mehner}, {Molaro}, {Nunes}, {Pepe}, {Santos}, \& {Sozzetti}}]{seidel:2023}
{Seidel}, J.~V., {Borsa}, F., {Pino}, L., {et~al.} 2023, \aap, 673, A125, \dodoi{10.1051/0004-6361/202245800}

\bibitem[{{Showman} {et~al.}(2009){Showman}, {Fortney}, {Lian}, {Marley}, {Freedman}, {Knutson}, \& {Charbonneau}}]{showman:2009}
{Showman}, A.~P., {Fortney}, J.~J., {Lian}, Y., {et~al.} 2009, \apj, 699, 564, \dodoi{10.1088/0004-637X/699/1/564}

\bibitem[{{Showman} \& {Polvani}(2011)}]{showman:2011}
{Showman}, A.~P., \& {Polvani}, L.~M. 2011, \apj, 738, 71, \dodoi{10.1088/0004-637X/738/1/71}

\bibitem[{{Sing} {et~al.}(2019){Sing}, {Lavvas}, {Ballester}, {Lecavelier des Etangs}, {Marley}, {Nikolov}, {Ben-Jaffel}, {Bourrier}, {Buchhave}, {Deming}, {Ehrenreich}, {Mikal-Evans}, {Kataria}, {Lewis}, {L{\'o}pez-Morales}, {Garc{\'\i}a Mu{\~n}oz}, {Henry}, {Sanz-Forcada}, {Spake}, {Wakeford}, \& {The PanCET collaboration}}]{sing:2019}
{Sing}, D.~K., {Lavvas}, P., {Ballester}, G.~E., {et~al.} 2019, \aj, 158, 91, \dodoi{10.3847/1538-3881/ab2986}

\bibitem[{{Siverd} {et~al.}(2012){Siverd}, {Beatty}, {Pepper}, {Eastman}, {Collins}, {Bieryla}, {Latham}, {Buchhave}, {Jensen}, {Crepp}, {Street}, {Stassun}, {Gaudi}, {Berlind}, {Calkins}, {DePoy}, {Esquerdo}, {Fulton}, {F{\H u}r{\'e}sz}, {Geary}, {Gould}, {Hebb}, {Kielkopf}, {Marshall}, {Pogge}, {Stanek}, {Stefanik}, {Szentgyorgyi}, {Trueblood}, {Trueblood}, {Stutz}, \& {van Saders}}]{siverd:2012}
{Siverd}, R.~J., {Beatty}, T.~G., {Pepper}, J., {et~al.} 2012, \apj, 761, 123, \dodoi{10.1088/0004-637X/761/2/123}

\bibitem[{{Tan} \& {Komacek}(2019)}]{tan:2019b}
{Tan}, X., \& {Komacek}, T.~D. 2019, \apj, 886, 26, \dodoi{10.3847/1538-4357/ab4a76}

\bibitem[{{Tan} \& {Showman}(2020)}]{tan:2020}
{Tan}, X., \& {Showman}, A.~P. 2020, \apj, 902, 27, \dodoi{10.3847/1538-4357/abb3d4}

\bibitem[{{Taylor} {et~al.}(2020){Taylor}, {Parmentier}, {Irwin}, {Aigrain}, {Lee}, \& {Krissansen-Totton}}]{taylor:2020}
{Taylor}, J., {Parmentier}, V., {Irwin}, P. G.~J., {et~al.} 2020, \mnras, 493, 4342, \dodoi{10.1093/mnras/staa552}

\bibitem[{{Thorngren} {et~al.}(2019){Thorngren}, {Gao}, \& {Fortney}}]{thorngren:2019}
{Thorngren}, D., {Gao}, P., \& {Fortney}, J.~J. 2019, \apjl, 884, L6, \dodoi{10.3847/2041-8213/ab43d0}

\bibitem[{{Tinetti} {et~al.}(2012){Tinetti}, {Beaulieu}, {Henning}, {Meyer}, {Micela}, {Ribas}, {Stam}, {Swain}, {Krause}, {Ollivier}, {Pace}, {Swinyard}, {Aylward}, {van Boekel}, {Coradini}, {Encrenaz}, {Snellen}, {Zapatero-Osorio}, {Bouwman}, {Cho}, {Coud{\'e} de Foresto}, {Guillot}, {Lopez-Morales}, {Mueller-Wodarg}, {Palle}, {Selsis}, {Sozzetti}, {Ade}, {Achilleos}, {Adriani}, {Agnor}, {Afonso}, {Prieto}, {Bakos}, {Barber}, {Barlow}, {Batista}, {Bernath}, {B{\'e}zard}, {Bord{\'e}}, {Brown}, {Cassan}, {Cavarroc}, {Ciaravella}, {Cockell}, {Coustenis}, {Danielski}, {Decin}, {Kok}, {Demangeon}, {Deroo}, {Doel}, {Drossart}, {Fletcher}, {Focardi}, {Forget}, {Fossey}, {Fouqu{\'e}}, {Frith}, {Galand}, {Gaulme}, {Hern{\'a}ndez}, {Grasset}, {Grassi}, {Grenfell}, {Griffin}, {Griffith}, {Gr{\"o}zinger}, {Guedel}, {Guio}, {Hainaut}, {Hargreaves}, {Hauschildt}, {Heng}, {Heyrovsky}, {Hueso}, {Irwin}, {Kaltenegger}, {Kervella}, {Kipping}, {Koskinen}, {Kov{\'a}cs}, {La Barbera}, {Lammer}, {Lellouch}, {Leto}, {Lopez
  Morales}, {Lopez Valverde}, {Lopez-Puertas}, {Lovis}, {Maggio}, {Maillard}, {Maldonado Prado}, {Marquette}, {Martin-Torres}, {Maxted}, {Miller}, {Molinari}, {Montes}, {Moro-Martin}, {Moses}, {Mousis}, {Nguyen Tuong}, {Nelson}, {Orton}, {Pantin}, {Pascale}, {Pezzuto}, {Pinfield}, {Poretti}, {Prinja}, {Prisinzano}, {Rees}, {Reiners}, {Samuel}, {S{\'a}nchez-Lavega}, {Forcada}, {Sasselov}, {Savini}, {Sicardy}, {Smith}, {Stixrude}, {Strazzulla}, {Tennyson}, {Tessenyi}, {Vasisht}, {Vinatier}, {Viti}, {Waldmann}, {White}, {Widemann}, {Wordsworth}, {Yelle}, {Yung}, \& {Yurchenko}}]{tinetti:2012}
{Tinetti}, G., {Beaulieu}, J.~P., {Henning}, T., {et~al.} 2012, Experimental Astronomy, 34, 311, \dodoi{10.1007/s10686-012-9303-4}

\bibitem[{{Wong} {et~al.}(2021){Wong}, {Shporer}, {Zhou}, {Kitzmann}, {Komacek}, {Tan}, {Tronsgaard}, {Buchhave}, {Vissapragada}, {Greklek-McKeon}, {Rodriguez}, {Ahlers}, {Quinn}, {Furlan}, {Howell}, {Bieryla}, {Heng}, {Knutson}, {Collins}, {McLeod}, {Berlind}, {Brown}, {Calkins}, {de Leon}, {Esparza-Borges}, {Esquerdo}, {Fukui}, {Gan}, {Girardin}, {Gnilka}, {Ikoma}, {Jensen}, {Kielkopf}, {Kodama}, {Kurita}, {Lester}, {Lewin}, {Marino}, {Murgas}, {Narita}, {Pall{\'e}}, {Schwarz}, {Stassun}, {Tamura}, {Watanabe}, {Benneke}, {Ricker}, {Latham}, {Vanderspek}, {Seager}, {Winn}, {Jenkins}, {Caldwell}, {Fong}, {Huang}, {Mireles}, {Schlieder}, {Shiao}, \& {Noel Villase{\~n}or}}]{wong:2021:2109}
{Wong}, I., {Shporer}, A., {Zhou}, G., {et~al.} 2021, \aj, 162, 256, \dodoi{10.3847/1538-3881/ac26bd}

\bibitem[{Zhou {et~al.}(2017)Zhou, Apai, Lew, \& Schneider}]{zhou:2017}
Zhou, Y., Apai, D., Lew, B. W.~P., \& Schneider, G. 2017

\bibitem[{{Zhou} {et~al.}(2022){Zhou}, {Apai}, {Tan}, {Lothringer}, {Lew}, {Casewell}, {Parmentier}, {Marley}, {Xu}, \& {Mayorga}}]{zhou:2022}
{Zhou}, Y., {Apai}, D., {Tan}, X., {et~al.} 2022, \aj, 163, 17, \dodoi{10.3847/1538-3881/ac3095}

\end{thebibliography}




\end{document}